\documentclass[letterpaper,11pt]{article}
\pdfoutput=1

\usepackage{hyperref}
\hypersetup{
    colorlinks=true,       
    linkcolor=darkcyan,          
    citecolor=darkcyan,        
    filecolor=darkcyan,      
    urlcolor=darkcyan           
}

\usepackage{float}
                     
\usepackage{braket,slashed,bm}
\usepackage{array,multirow}
\usepackage[normalem]{ulem}
\usepackage{xcolor,youngtab}
\usepackage{verbatim}

\usepackage[T1]{fontenc} 
\usepackage[font=small,labelfont=bf,tableposition=top]{caption}

\DeclareCaptionLabelFormat{andtable}{#1~#2  \&  \tablename~\thetable}

\usepackage{mathrsfs}
\usepackage{booktabs}
\usepackage{adjustbox}
\usepackage{mathtools}

\usepackage{soul}

\usepackage{tikz}
\usetikzlibrary{arrows,decorations.pathmorphing,backgrounds,positioning,fit,petri,automata,shadows,calendar,mindmap,
decorations.markings,calc}

\definecolor{labelkey}{rgb}{0,0.5,0.0}

\usepackage{subcaption}
\usepackage{xspace}
\usepackage{slashed}
\usepackage{array,multirow}
\usepackage{graphicx}
\usepackage{graphics}
\usepackage{epsfig}
\usepackage{amsfonts}
\usepackage{amsmath}
\usepackage{amssymb}
\usepackage{dsfont}
\usepackage{color}
\usepackage{xcolor}
\usepackage{cite}
\usepackage{pdflscape}
\usepackage{pifont}
\usepackage{pgfplots}
\usepackage{tikz} 
\usepackage{pgfplotstable}
\usepackage{bbold}
\usepackage[makeroom]{cancel}

\newcommand{\vvbb}{$2\nu\beta\beta$ }

\usepackage{bm}  %

\newcommand{\beq}{\begin{equation}}
\newcommand{\eeq}{\end{equation}}
\newcommand{\be}{\begin{equation}}
\newcommand{\ee}{\end{equation}}
\newcommand{\bea}{\begin{eqnarray}}
\newcommand{\eea}{\end{eqnarray}}
\newcommand{\ben}{\begin{eqnarray*}}
\newcommand{\een}{\end{eqnarray*}}

\newcommand{\bma}{\begin{pmatrix}}
\newcommand{\ema}{\end{pmatrix}}

\def\lixo#1{}

\def\slashchar#1{\setbox0=\hbox{$#1$}           
  \dimen0=\wd0                                    
  \setbox1=\hbox{/} \dimen1=\wd1                  
  \ifdim\dimen0>\dimen1                           
    \rlap{\hbox to \dimen0{\hfil/\hfil}}            
    #1                                             
  \else                                          
    \rlap{\hbox to \dimen1{\hfil$#1$\hfil}}        
    /                                           
 \fi}                                           %

\newcommand{\dslash}[1]{#1 \llap{/\kern-0.5pt}}
\newcommand{\Dslash}[1]{#1 \llap{/\kern+1.5pt}}
\newcommand{\DDslash}[1]{#1 \llap{/\kern+2.3pt}}
\newcommand{\dslashh}[1]{#1 \llap{/\kern+1pt}}

\definecolor{cadmiumgreen}{rgb}{0.0, 0.42, 0.24}
\definecolor{darkpastelgreen}{rgb}{0.01, 0.75, 0.24}
\definecolor{darkspringgreen}{rgb}{0.09, 0.45, 0.27}
\definecolor{forestgreen(web)}{rgb}{0.13, 0.55, 0.13}
\definecolor{forestgreen(traditional)}{rgb}{0.0, 0.27, 0.13}
\definecolor{cobalt}{rgb}{0.0, 0.28, 0.67}
\definecolor{darkblue}{rgb}{0.0, 0.0, 0.75}
\definecolor{darkred}{rgb}{0.55, 0.0, 0.0}
\definecolor{palatinatepurple}{rgb}{0.41, 0.16, 0.38}
\definecolor{burntorange}{rgb}{0.8, 0.33, 0.0}
\definecolor{darkcyan}{rgb}{0.0, 0.666666, 0.666666}

\begin{document}

\begin{titlepage}

\begin{flushright}
 LA-UR-24-33136, INT-PUB-24-062
\\
\end{flushright}

\vspace{1.cm}

\begin{center}
{\LARGE  \bf 
$2\nu\beta\beta$   Spectrum  
in Chiral Effective Field Theory
}

\vspace{2cm}
{\large \bf  Saad el Morabit$^{a,b,}$\footnote{\href{mailto:s.elmorabit2@uva.nl}{s.elmorabit2@uva.nl}}, 
Ryan Bouabid$^{c,}$\footnote{\href{mailto:ryan.bouabid@duke.edu}{ryan.bouabid@duke.edu}}, 
Vincenzo Cirigliano$^{d,}$\footnote{\href{mailto:cirigv@uw.edu}{cirigv@uw.edu}}, \\ 
Jordy de Vries$^{a,b,}$\footnote{\href{mailto:j.devries4@uva.nl}{j.devries4@uva.nl}}, 
Luk\'{a}\v{s} Gr\'{a}f$^{\, b,e,}$\footnote{\href{mailto:lukas.graf@nikhef.nl}{lukas.graf@nikhef.nl}}, 
Emanuele Mereghetti$^{f,}$\footnote{\href{mailto:emereghetti@lanl.gov}{emereghetti@lanl.gov}} } 
\vspace{0.5cm}

{\large 
$^a$ 
{\it 
Institute for Theoretical Physics Amsterdam, University of Amsterdam, Science Park 904, 1098 XH Amsterdam, The Netherlands}}

\vspace{0.25cm}
{\large 
$^b$ 
{\it 
Nikhef, Theory Group, Science Park 105, 1098 XG, Amsterdam, The Netherlands}}

\vspace{0.25cm}
{\large 
$^c$ 
{\it 
Department of Physics, Duke University, Durham, NC, 27708, USA}}

\vspace{0.25cm}
{\large 
$^d$ 
{\it 
Institute for Nuclear Theory, University of Washington, Seattle WA 98195-1550, USA}}

\vspace{0.25cm}
{\large 
$^e$ 
{\it 
Institute of Particle and Nuclear Physics, Faculty of Mathematics and Physics, Charles University in Prague, V Hole\v{s}ovi\v{c}k\'ach 2, 180 00 Praha 8, Czech Republic}}

\vspace{0.25cm}
{\large 
$^f$ 
{\it Theoretical Division, Los Alamos National Laboratory,
Los Alamos, NM 87545, USA}}

\end{center}

\vspace{0.2cm}

\begin{abstract}
\vspace{0.1cm}

We investigate 
two-neutrino double beta decay ($2\nu\beta\beta$) 
in chiral effective field theory. 
We find contributions from weak magnetism and double-weak pion-exchange  
at next-to-leading-order in the chiral power counting. We 
discuss the impact 
of the chiral corrections 
on the electron spectra and find that 
they 
should be included in 
analyses of $2\nu\beta\beta$ decay 
that aim to uncover new physics signatures in the electron spectrum. 
 We illustrate this point by  revisiting the effect of sterile neutrinos and non-standard charged interactions.
We also find that the pion-exchange 
contributions 
involve 
nuclear matrix elements that 
are related to 
those appearing in neutrinoless double beta decay ($0\nu \beta \beta$). We investigate whether 
the $0\nu \beta \beta$ nuclear matrix elements can be obtained from detailed measurements of the energy spectrum of the outgoing electrons in $2\nu\beta\beta$ transitions.

\end{abstract}

\vfill
\end{titlepage}

\section{Introduction}

Double-weak processes, where two nucleons undergo $\beta$-decay and produce two (anti)neutrinos and two electrons $\left(2\nu\beta\beta\right)$, are the slowest processes ever measured~\cite{Barabash:2023dwc, XENON:2019dti}.
For example, the lifetime of ${}^{76}$Ge is $~10^{21}$ yr, a staggering eleven orders of magnitude longer than the age of the universe \cite{GERDA:2023wbr}. 
A counterpart to $2\nu\beta\beta$ is the hypothetical \emph{neutrinoless} double beta decay $(0\nu\beta\beta)$, which involves the same transition, but without final-state neutrinos. The detection of $0\nu\beta\beta$ is one of the major goals 
for nuclear and particle physics in the coming decades~\cite{Agostini:2022zub,Adams:2022jwx}. A signal of the process where two electrons are created without accompanying anti-neutrinos would prove that lepton number is not conserved in nature and demonstrate that neutrinos, unlike any other fermions in the Standard Model (SM), are Majorana particles. The most stringent limit on $0\nu\beta\beta$, 
$T^{0\nu}_{1/2} (^{136}{\rm Xe}) > 3.8\cdot 10^{26} \, \mathrm{yr}$ (90\% C.L.), was established earlier this year by the KamLAND-Zen collaboration \cite{KamLAND-Zen:2024eml}, and next-generation experiments 
will have sensitivity to half-lives in excess of $10^{28}\, \mathrm{yr}$~\cite{Adams:2022jwx}.

The development of modern $0\nu\beta\beta$ and dark-matter direct detection experiments has led to experiments with very large masses and extended data collection periods. New $0\nu\beta\beta$ and direct dark matter detection experiments will measure an unprecedented number of $2\nu\beta\beta$ events  with very low background
and can thus measure the  electron energy spectrum with high accuracy \cite{Adams:2022jwx,CUPID:2019imh,nEXO:2018ylp,LEGEND:2021bnm, XENON:2022evz}.
While $2\nu\beta\beta$ forms a background for $0\nu\beta\beta$, the $2\nu\beta\beta$ electron spectrum has been used 
both 
as a test for 
nuclear structure calculations~\cite{Simkovic:2018rdz,KamLAND-Zen:2019imh} 
as well as a tool 
to probe physics beyond the Standard Model (BSM)~\cite{Doi:1987rx,Burgess:1992dt,Burgess:1993xh,Blum:2018ljv,Deppisch:2020mxv,Deppisch:2020sqh,Bolton:2020ncv,Agostini:2020cpz,Nitescu:2020xlr,Agostini:2015nwa,CUPID:2024qnd,GERDA:2022ffe,Kharusi:2021jez,CUPID-0:2022yws,Diaz:2013ywa,Diaz:2013saa}.
Using the $2\nu\beta\beta$ measurements in such a way requires an accurate theoretical description of the electron spectrum. A lot of effort has gone into electromagnetic corrections to the spectrum \cite{Doi:1985dx,Kotila:2012zza,Nitescu:2024tvj} and the effect of the outgoing lepton energy on the nuclear matrix elements \cite{Simkovic:2018rdz}. 

In this work, we focus on a different type of correction associated with the form of the weak currents and 
more generally weak transition operators. The use of chiral EFT ($\chi$EFT) \cite{Weinberg:1991um}, the low-energy EFT of QCD that describes interactions of nucleons and pions, has proven very valuable in the description of $0\nu\beta\beta$ decay 
\cite{Cirigliano:2018hja,Wirth:2021pij,Belley:2023lec}. 
Here we apply the EFT framework to $2\nu\beta\beta$ decay
and use $\chi$EFT to organize corrections to the $2\nu\beta\beta$ spectrum in a systematic expansion in the ratio of scales involved in the problem.
We find two main corrections that should be included in the analysis of $2\nu\beta\beta$ spectra in order to isolate BSM physics:
\begin{itemize}
\item The first correction arises from the subleading 
one-nucleon 
contributions associated with weak magnetism  (see related work in Ref.~\cite{Doi:1985dx,Barbero:1998jh}). These are the first corrections in a multipole expansion of the weak currents, and affect the spectral shape at the per-mille level.

\item A second type of correction arises from two-nucleon weak currents induced by  pion exchange, 
along with the associated short-range two-nucleon operators needed to renormalize the amplitudes. 
We will show that these pion-induced  chiral corrections to the $2\nu\beta\beta$ electron spectrum depend on nuclear matrix elements (NMEs) that also appear in the description of $0\nu\beta\beta$ induced by massive Majorana neutrino exchange. 
\end{itemize}

A key problem in interpreting the results of $0\nu\beta\beta$ searches 
(null so far)  
is 
the theoretical prediction of the NMEs that suffer from an uncertainty up to an order of magnitude. Recent developments using $\chi$EFT and \emph{ab initio} nuclear many-body calculations have been used to reduce this uncertainty, but a more direct approach to the NMEs would be very welcome.
In this light, there have been attempts to find correlations between $0\nu\beta\beta$ NMEs and other observables like ion double-charge exchange \cite{Amos:2007rb,Santopinto:2018nyt,NUMEN:2022ton}, double Gamow-Teller matrix elements
\cite{Shimizu:2017qcy,Yao:2022usd,Jokiniemi:2023bes}, nucleon-nucleon phase shifts \cite{Belley:2024zvt}, or double-magnetic dipole transitions
\cite{Romeo:2021zrn,Romeo:2024xkm}. Although these studies are important and might reveal deficiencies in the many-body methods used to compute nuclear wavefunctions and matrix elements, these correlations are determined mainly phenomenologically and lack a precise theoretical understanding, i.e., in terms of matrix elements of the same two-body operator appearing in $0\nu\beta\beta$. 
The $\chi$EFT or low-energy QCD relation we find here is much more concrete and has a firm theoretical basis.

This paper is organized as follows.
In Section \ref{sec:2nu_amplitude} we derive the $2\nu\beta\beta$ amplitude, including higher-order corrections in the chiral counting. To separate phase-space integrations from NMEs we discuss the lepton energy expansion introduced in Ref. \cite{Simkovic:2018rdz}, in Section~\ref{sec:expansion}, and verify it using different nuclear methods. We discuss  how chiral corrections affect the $2\nu\beta\beta$ spectrum and the role of nuclear uncertainties in Section~\ref{sec:sensitivity}. We compare the chiral corrections identified in this work to representative BSM corrections to the spectrum in Section~\ref{sec:Beyond-the-Standard-Model Contributions to 2nu}. We summarize, conclude, and present an outlook in Section~\ref{sec:conclusions}.

\section{The $2\nu\beta\beta$ Amplitude and Differential Decay rate}\label{sec:2nu_amplitude}
\begin{figure}[t]
\begin{center}
\includegraphics[scale =0.7]{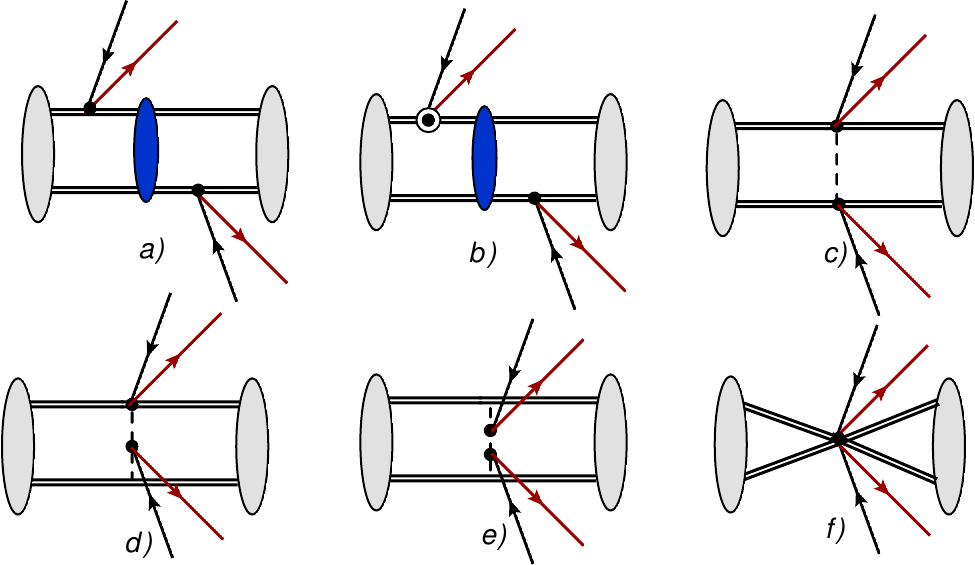}
\end{center}
\caption{Diagrams contributing to $2\nu\beta\beta$. Gray ellipses denote the initial- and final-state nuclear wave functions, whereas the blue ellipse denotes strong interactions among intermediate nucleons that generate nuclear excited states. The LO diagram (a) describes two nucleons (double lines) undergoing $\beta$-decay into an electron (red lines) and anti-neutrinos (single black lines) through LO weak vertices (black dots). Diagram (b) involves a subleading $\beta$-decay vertex from weak magnetism. Diagrams (c)-(e) correspond to pion-exchange double-weak currents (pions are denoted by dashed lines). Finally diagram (f) describes a double-weak contact term that renormalizes diagrams (c)-(e).}
\label{2nubb_LONLO}
\end{figure}

We consider the process \begin{equation}
0^+(E_i) \rightarrow 0^+(E_f) + e^-(E_{e1})+ e^-(E_{e2})+ \bar \nu_e (E_{\nu1}) + \bar \nu_e (E_{\nu2})\,,
\end{equation}
where $0^+(E_i)$ and $0^+(E_f)$ denote the initial- and final-state atomic ground states with vanishing total angular momentum and positive parity. The dominant contribution to this decay arises from diagram (a) in Fig.~\ref{2nubb_LONLO}. This contribution is very well studied in the literature and its amplitude can be written as \cite{Doi:1985dx,Haxton:1984ggj,Simkovic:2018rdz}
\begin{eqnarray}\label{TLO}
T_{\beta\times\beta} &=& -\frac{4 g_A^2 G_F^2 V_{ud}^2}{3 m_e}\,\bigg(  M_{GT}^K L_{11}^i L_{22}^i -M_{GT}^L L_{12}^i L_{21}^i \bigg)\,\,
\end{eqnarray}
in terms of the Fermi constant $G_F$, the nucleon axial charge $g_A \simeq 1.27$, and the CKM element $V_{ud} = 0.974$. We define the lepton current $L^\mu_{ij} = \bar e_L(p_{e,i}) \gamma^\mu \nu_L(p_{\nu,j})$ and the dimensionless structures
\begin{equation}\label{eq:LOME}
M^{K,L}_{GT} = m_e \sum_n G_n \frac{E_n - \frac{1}{2} (E_i + E_f)}{[E_n - \frac{1}{2} (E_i + E_f)]^2 - \epsilon^2_{K,L}}\,,
\end{equation}
where we encounter the nuclear matrix element associated to a double Gamow-Teller (GT) transition
\begin{eqnarray}\label{eq:GnLO}
G_n = \langle f| \sum_k  \vec \sigma_k\tau_k^+ |n\rangle \cdot \langle n | \sum_l \vec \sigma_l \tau_l^+ | i\rangle\,,
\end{eqnarray}
where $k,l$ sum over the nucleons inside the nucleus and $n$ denotes an intermediate nuclear state with quantum number $J^\pi = 1^+$ and energy $E_n$. In addition, we introduced the combination of lepton energies
\begin{equation}
\epsilon_K = (E_{22} - E_{11})/2\,,\qquad \epsilon_L =  (E_{12} - E_{21})/2\,,
\end{equation}
in terms of $E_{ij} = E_{ei} + E_{\nu j}$. We have neglected a double Fermi NME in Eq.~\eqref{TLO} because it is suppressed by isospin breaking.

Eqs. \eqref{TLO}, \eqref{eq:LOME}
and \eqref{eq:GnLO} are nothing but the leading contributions in a multipole and chiral expansions,
which can be organized in a double expansion in the small parameters $\epsilon_\pi$ and $\epsilon_\chi$, with
\begin{equation}
\epsilon_\chi = \mathcal O\left(\frac{k_F}{\Lambda_\chi} \right),    \qquad \epsilon_\pi = \mathcal O\left(\frac{\mathcal Q}{k_F}\right),
\end{equation}
where here by $k_F$ we denote scales of the order of the typical momentum of nucleons inside nuclei,
$k_F \sim m_\pi \sim 100$ MeV,
while $\mathcal Q$ denote the reaction's $\mathcal Q$-value, $\mathcal Q \sim E_{1, 2} \sim m_e \sim $ few MeVs.
We can roughly divide higher order contributions into 2 classes. In the first class we encounter diagram \ref{2nubb_LONLO}(b)  with an intermediate nuclear state such as in diagram \ref{2nubb_LONLO}(a) but with subleading weak currents or with higher order terms in the lepton momenta. 
The nuclear matrix element in Eq. \eqref{eq:LOME} can be  replaced by the more general expression 
\begin{eqnarray}\label{eq:NLOME}
M^K = \frac{m_e}{g_A^2} \sum_n  \Bigg( \frac{1}{E_n - \frac{1}{2} (E_i + E_f) - \epsilon_{K}}\,
\langle f| \sum_k  \vec J_{V-A}(\vec q_{11}) |n\rangle \cdot \langle n | \vec J_{V-A}(\vec q_{22}) | i\rangle \nonumber  \\
+ \frac{1}{E_n - \frac{1}{2} (E_i + E_f) + \epsilon_{K}}\,
\langle f| \sum_k  \vec J_{V-A}(\vec q_{22}) |n\rangle \cdot \langle n | \vec J_{V-A}(\vec q_{11}) | i\rangle \Bigg),
\end{eqnarray}
with $\vec q_{ij} = \vec p_{ei} + \vec p_{\nu j}$, and $M^L$ can be obtained
by replacing $\epsilon_K \rightarrow \epsilon_L$, 
$\vec q_{11} \rightarrow \vec q_{12}$
and $\vec q_{22} \rightarrow \vec q_{21}$.
At zero lepton momentum,
the leading corrections come from two-body axial currents. These scale as
$    \mathcal O\left(\epsilon_\chi^2\right)$,
but amount to a shift in the total rate and do not affect the energy spectrum.
At non-zero lepton momentum,
the nuclear matrix elements
in Eq. \eqref{eq:NLOME}
can be expanded by using the multipole expansion \cite{Donnelly:1975ze,Walecka:1995mi}. This yields an expansion in $\mathcal Q/k_F$ or $\mathcal Q/m_N$, with $m_N$ the nucleon mass, and in general affects both the total rate and the shape of the electron energy distribution.
The leading contribution in the multipole expansion arises from weak magnetism (WM), 
at $\mathcal O(\mathcal Q/\Lambda_\chi) \sim 10^{-3}$. At $\mathcal O(\mathcal Q^2/k_F^2) \sim 10^{-4}$ one would find contributions from the induced pseudoscalar form factor, from the radii of the axial form factor, etc.
We will focus here on the contribution from WM which is introduced through perturbing the nuclear current with a vectorial term 
proportional to $g_M$
\begin{equation}
    J^\mu_N = g_V v^\mu - 2g_AS^\mu + \left[\frac{i g_M}{m_N}\epsilon^{\mu \nu \alpha\beta}v_\alpha S_\beta q_\nu\right]\;, 
\end{equation}
where $g_M=1+\kappa_1=4.7$ is the magnetic coupling, $v^\mu = \left(1, \vec{0}\right)$ and  $S^\mu = \left(0, \frac{1}{2}\vec{\sigma}\right)$ are the nucleon velocity and spin.

In the second class of corrections we encounter diagrams \ref{2nubb_LONLO}(c)-(e) which involve an irreducible double-$\beta$ two-nucleon transition.
By power counting these diagrams, we can see that they are suppressed by
$\mathcal O(\mathcal Q/\Lambda_\chi)$ with respect to the leading-order contributions.
As for WM, these corrections do affect the electron spectrum.
As the typical pion momentum $|\vec q^{\,}|$ is of order $k_F$, and much larger than the electron and neutrino energies, these diagrams generate ``neutrino potentials'', analogous to those found in $0\nu\beta\beta$. Their contribution to $2\nu\beta\beta$ only depends on the initial and final state wavefunctions and not on the intermediate nuclear states. As a consequence, the amplitude does not contain the energy denominators in Eqs. \eqref{eq:LOME} and \eqref{eq:NLOME}, and gives rise to a different dependence on the electron energy compared to the LO diagrams.

Finally, we encounter diagram \ref{2nubb_LONLO}(f) that involves a double-weak contact interaction between four-nucleons and four-leptons that is necessary to renormalize diagrams \ref{2nubb_LONLO}(c)-(e) and, similarly, do affect the electron spectrum.

We now turn to the actual computation. The sum of irreducible diagrams (c)-(e) in Fig.~\ref{2nubb_LONLO} contribute to the double-weak two-nucleon transition operator (we will call it the two-neutrino potential) which can be computed using the LO $\chi$EFT Lagrangian (see e.g.~Ref.~\cite{Bernard:1995dp}). In momentum space the two-neutrino potential from diagrams (c)-(e) reads
\begin{align}\label{eq:V2nu}
 &   V_{2\nu}(\vec q\,) = -\frac{(8 G_F^2 V_{ud}^2)}{2F_\pi^2}(\tau_1^+ \tau_2^+)  \frac{1}{\vec q^{\,2} + m_\pi^2} \left( L_{11}^\mu L_{22}^\nu - L_{12}^\mu L_{21}^\nu \right)
    \Bigg\{ g_V^2 \delta^{\mu 0}\delta^{\nu 0}  + \frac{g_A^2}{3} \delta^{\mu i} \delta^{\nu i}  \nonumber \\
&
\times \left[ \vec\sigma_1 \cdot \vec \sigma_2 \left(1 + \frac{4}{3} \frac{\vec q^{\, 2}}{\vec q^{\, 2} + m_\pi^2} + \frac{4}{3}\left(\frac{\vec q^{\, 2}}{\vec q^{\, 2} + m_\pi^2}\right)^2 \right) - \frac{4}{3}S_{12} \frac{\vec q^{\, 2}}{\vec q^{\, 2} + m_\pi^2} \left(1 + \frac{\vec q^{\, 2}}{\vec q^{\, 2} + m_\pi^2}\right)  \right]
\Bigg    \}
\end{align}
in terms of $\vec q$, the momentum transfer between nucleon $1$ and $2$ and we have neglected the lepton momenta.
The tensor operator is defined as $S_{12} = \vec\sigma_1 \cdot \vec \sigma_2 - 3 \vec\sigma_1 \cdot \vec q\,  \vec \sigma_2 \cdot \vec q/{\vec q^{\, 2}}$.
The two-neutrino potentials are of pion range and have a Coulomb-like scaling for large virtual momenta $|\vec q^{\,}| \gg m_\pi$. 
Their matrix elements on leading-order chiral wavefunctions therefore depend logarithmically on the ultraviolet cut-off used to regulate the short-range nuclear force. Similar divergences were observed in the case of the pion exchange potential \cite{Kaplan:1996xu},
of pion-nucleus scattering operators
\cite{Borasoy:2001gq,Cirigliano:2024ocg} and for the leading-order neutrino potential in $0\nu\beta\beta$\cite{Cirigliano:2018hja,Cirigliano:2019vdj}, and can be cured by including short-range $2\nu\beta\beta$ operators at the same order as the contributions in Eq. \eqref{eq:V2nu}.
These will yield a short-range potential
\begin{align}\label{eq:V2nuShort}
 &   V^{S}_{2\nu}(\vec q\,) = -\frac{(8 G_F^2 V_{ud}^2)}{2F_\pi^4}(\tau_1^+ \tau_2^+)  \left( L_{11}^\mu L_{22}^\nu - L_{12}^\mu L_{21}^\nu \right) \left( g^{\rm NN}_{2\nu, \rm F}\, \delta^{\mu 0}\delta^{\nu 0}  - g^{\rm NN}_{2\nu, \rm GT} \, \delta^{\mu i} \delta^{\nu i} \right),
\end{align}
with 
\begin{equation}
    g^{\rm NN}_{2\nu, \rm F} \sim g^{\rm NN}_{2\nu, \rm GT} = \mathcal O(1). 
\end{equation}
The Gamow-Teller low-energy constant (LEC), $g^{\rm NN}_{2\nu, \rm GT}$, is related to the contact contribution to the isotensor axial polarizability introduced in Refs. \cite{Shanahan:2017bgi,Tiburzi:2017iux}.
We refrain here from writing down the complete set of chiral invariant operators that lead to Eq. \eqref{eq:V2nuShort}, but notice that chiral symmetry  relates some of them to the pion-nucleus scattering operators introduced in Ref. \cite{Cirigliano:2024ocg}.

The two-neutrino potentials contribute to the Fourier-transformed amplitude through
\begin{eqnarray}\label{TLO2}
T_{\beta\beta} &=& \langle f| \sum_{m,n} \int \frac{d^3 q}{(2\pi)^3} e^{i \vec q \cdot \vec r} V_{2\nu}(\vec q) | i \rangle\,,
\end{eqnarray}
where $m,n$ sums over all nucleons in the nucleus. We write this more explicitly as
\begin{eqnarray}\label{Tbb}
T_{\beta\beta} &=& -\frac{(8 G_F^2 V_{ud}^2)}{m_e}\left[ \epsilon_F \left( L_{11}^0 L_{22}^0 - L_{12}^0 L_{21}^0 \right) + \epsilon_{GT}  \left( L_{11}^i L_{22}^i - L_{12}^i L_{21}^i \right)\right]\,,
\end{eqnarray}
where
\begin{eqnarray}
\epsilon_F &=& \frac{m_e g_V^2}{2 F_\pi^2} \frac{1}{4\pi R_A} \left(M_F(m_\pi) + \frac{m_\pi^2}{F_\pi^2} g_{2\nu, \rm F}^{\rm NN} \,  M_{F,\, sd} \right)\,\\
\epsilon_{GT} &=& \frac{m_e g_A^2}{6 F_\pi^2} \frac{1}{4\pi R_A}  \left( M^{AA}_{GT}(m_\pi)-2 M^{AP}_{GT}(m_\pi)+4 M^{PP}_{GT}(m_\pi) - \frac{3 m_\pi^2}{ g_A^2 F_\pi^2} g_{2\nu, \rm GT}^{\rm NN} \,  M_{F,\, sd}  \right)\,.
\end{eqnarray}

The nuclear matrix elements are defined through
\begin{eqnarray}
M_{F, (sd)}(m_\pi) &=& \langle 0^+| \sum_{m,n} h_{F, (sd)}(r,m_\pi) \tau^+_m  \tau^+_n | 0^+\rangle\,, \nonumber\\
M^{ab}_{GT}(m_\pi) &=& \langle 0^+| \sum_{m,n} h^{ab}_{GT}(r,m_\pi) \vec \sigma_m \cdot \vec \sigma_n \tau^+_m  \tau^+_n | 0^+\rangle\,, 
\end{eqnarray}
where $ab =\{AA, AP,PP \}$ and the radial functions are given by\footnote{The integral in Eq. \eqref{eq:radialFermiSD} is formally divergent, and should be interpreted as a regularization of the $\delta^{(3)}(r)$ function. Eq. \eqref{eq:radialFermiSD} can be regulated in many ways, e.g.\ by adding a vector form factor, see Eq. (30) in Ref.~\cite{Cirigliano:2017djv}.
}  
\begin{eqnarray}
h_F(r,m_\pi) &=& \frac{2 R_A}{\pi} \int_0^{\infty} dq \frac{q^2}{q^2+m_\pi^2} h_F(q^2) \frac{\sin qr}{qr}\,, \label{eq:radialFermi}\\
h_{F, sd}(r) &=& \frac{2 R_A}{\pi m_\pi^2} \int_0^{\infty} dq q^2 h_F(q^2) \frac{\sin qr}{qr}\,,
\label{eq:radialFermiSD}\\
h^{ab}_{GT}(r,m_\pi) &=& \frac{2 R_A}{\pi} \int_0^{\infty} dq \frac{q^2}{q^2+m_\pi^2} h^{ab}_{GT}(q^2) \frac{\sin qr}{qr}\,,
\end{eqnarray}
in terms of 
\begin{eqnarray}
h_F(q^2) = h^{AA}_{GT}(q^2)=  1\,,\qquad h^{AP}_{GT}(q^2) = -\frac{2}{3} \frac{q^2}{q^2+m_\pi^2}\,,\qquad h^{PP}_{GT}(q^2) = \frac{1}{3} \frac{q^4}{(q^2+m_\pi^2)^2}\,.
\end{eqnarray}
By power counting we observe that $\epsilon_F \sim \epsilon_{GT} = \mathcal O(m_e/(4\pi F_\pi)) = \mathcal O(m_e/\Lambda_\chi) \simeq 0.1\%$. 
In Eq. \eqref{Tbb} we neglect tensor operators, whose matrix elements are typically significantly  smaller than those of Fermi and Gamow-Teller operators \cite{Barea:2015kwa,Hyvarinen:2015bda,Menendez:2017fdf}.

The NMEs entering $\epsilon_{F,GT}$ very closely resemble those appearing in the description of $0\nu\beta\beta$. $0\nu\beta\beta$ induced by the exchange of light Majorana neutrinos involves $M_F(0)$ and $M^{ab}_{GT}(0)$. However, $0\nu\beta\beta$ induced by heavier sterile neutrinos does involve the required $M_F(m_\pi)$ and $M^{ab}_{GT}(m_\pi)$ and detailed calculations \cite{Blennow:2010th,Faessler:2014kka,Barea:2015zfa,Dekens:2023iyc,Dekens:2024hlz} of these NMEs have appeared. 
The ratios $M_F(m_\pi)/M_F(0)$ and $M^{ab}_{GT}(m_\pi)/M^{ab}_{GT}(0)$ are rather stable with respect to the applied nuclear many-body method  and show a much smaller spread than the values for $M_F(0)$ and $M^{ab}_{GT}(0)$ themselves as shown in Table~\ref{tab:comparison}, implying that an extraction of the NMEs evaluated at $m_\pi$ provides information about the $0\nu\beta\beta$ NMEs. 
 At the moment, given the unknown  low-energy constants  $g_{2\nu,\rm F}^{\rm NN}$ and $g_{2\nu,\mathrm{GT}}^{\rm NN}$, such a program cannot be carried out.
On a related point, 
since we do not control the LECs $g_{2\nu,\rm F}^{\rm NN}$ and $g_{2\nu,\mathrm{GT}}^{\rm NN}$ we will not include them in our numerical analysis and we can only provide an incomplete estimate the size of the pion-exchange corrections to $2 \nu \beta \beta$ observables.

The total amplitude at this order becomes
\begin{align}
T_{\beta\times\beta} + T_{\beta\beta}  &= -\frac{(8 G_F^2 V_{ud}^2)}{m_e}\,\bigg\{ \frac{g_A^2}{6} \left[ \left(\left(M_{GT}^K+ \frac{6 \epsilon_{GT}}{g_A^2}\right)\delta_{ij}+\frac{i g_M}{2 g_A m_N}\left(p_{11}-p_{22}\right)^k\epsilon_{ijk} M_{GT}^K\right) L_{11}^i L_{22}^j\right.\nonumber \\
&+ \left. \left(\left(M_{GT}^L + \frac{6  \epsilon_{GT}}{g_A^2}\right)\delta_{ij}+ \frac{i g_M}{2 g_A m_N}\left(p_{12}-p_{21}\right)^k\epsilon_{ijk}  M_{GT}^L \right) L_{12}^i L_{21}^j\right]\nonumber\\
&+\epsilon_F \left( L_{11}^0 L_{22}^0 - L_{12}^0 L_{21}^0 \right) \bigg\}\,,
\end{align}
where we introduce $p_{ij}=(E_{ij}, \vec{p}_{ij})$ to describe the change in nucleon four-momentum $q$ in terms of lepton four momenta.

\begin{table}[t]
\center
$\renewcommand{\arraystretch}{1.5}
\begin{array}{|l|cc|cc|}
\hline
 &  \multicolumn{2}{c|}{\text{}^{76} \text{Ge}} & \multicolumn{2}{c|}{\text{}^{136} \text{Xe}}\\
 \mathrm{NMEs} &  \mathrm{QRPA} &  \mathrm{Shell}\,\,\mathrm{Model} &  \mathrm{QRPA} &  \mathrm{Shell}\,\,\mathrm{Model}\\
 \hline
 M_{GT}^{(-1)} &0.069& 0.058&0.010&0.013\\
 \xi_{31} &0.10& 0.13 &0.39& 0.17\\
 \xi_{51} &0.017&0.022 &0.12&0.046\\
 \hline
 M_F(m_\pi) & -1.16  & -0.42 & -0.56 &-0.38\\
 M^{AA}_{GT}(m_\pi) & 3.57 & 1.91 & 2.03 & 1.55\\
 M^{AP}_{GT}(m_\pi) &-1.47 & -0.66 &-0.83 & -0.56 \\
 M^{PP}_{GT}(m_\pi) &0.50 &0.22 & 0.29 & 0.19\\
  M_{F,\, sd} &	-3.46&-1.46 &-1.53 &-1.28\\
  M_{F}(m_\pi) /M_{F}(0)  &	0.67&0.71 &0.63 &0.70\\
  M_{GT}^{AA}(m_\pi) /M_{GT}^{AA}(0)  &0.67&0.61 &0.64 &0.63\\
  M_{GT}^{AP}(m_\pi) /M_{GT}^{AP}(0)  &0.73&0.71 &0.70 &0.72\\
  M_{GT}^{PP}(m_\pi) /M_{GT}^{PP}(0)  &0.75&0.73 &0.73 &0.75\\
  M_{GT}(m_\pi) /M_{GT}(0)  &0.71&0.66 &0.68 &0.68\\\hline
 \Delta_0-1 &0.040& 0.020& 0.14& 0.077\\
 \Delta_2-1 &0.025 & 0.014 & 0.076 & 0.043 \\
 \hline
 \end{array}$
 
\caption{
Values of $2\nu\beta\beta$ and $0\nu\beta\beta$ NMEs computed in the quasi particle random phase approximation and nuclear shell model 
for $^{76}$Ge and $^{136}$Xe. 
The QRPA estimates of $M_{GT}^{-1}$, $\xi_{31}$
and $\xi_{51}$ are obtained from Ref. \cite{Simkovic:2018rdz}, while the NSM values are calculated using the matrix elements in Ref. \cite{Dekens:2024hlz}.
The $0\nu\beta\beta$ NMEs use the QRPA and NSM calculations of Refs. \cite{Hyvarinen:2015bda}
and \cite{Menendez:2017fdf}, respectively. The value of $M_{GT}^{(-1)}$ corresponds to no quenching of $g_A$.
}
\label{tab:comparison}
\end{table}

The corresponding decay rate, accounting for $1/(2!\times 2!)$ for identical electrons and neutrinos, is given by
\begin{eqnarray}
\Gamma &=& \frac{(G_F V_{ud})^4}{8 \pi^7 m_e^2} \int^{E_i-E_f-m_e}_{m_e} d E_{e1} \int^{E_i-E_f-E_{e1}}_{m_e} d E_{e2} \int^{E_i-E_f-E_{e1}-E_{e2}}_{0} d E_{\nu1}\nonumber\\
&&\times E_{e1} p_{e1} E_{e2} p_{e2} E_{\nu 1}^2 E_{\nu 2}^2 \times F(E_{e1},Z_f) \times F(E_{e2},Z_f)\times C_{2\nu}\,,\label{Gamma}
\end{eqnarray}
where we should interpret $E_{\nu 2} = (E_i - E_f  - E_{11} - E_{e2})$ and
\begin{align}\label{eq:C2nu}
 C_{2\nu}  = & \    \frac{g_A^4}{3}\bigg\{\left[  (M_{GT}^K)^2 +  (M_{GT}^L)^2 +  M_{GT}^K M_{GT}^L\right]\
 \nonumber \\
 \times & \left[1-\frac{2g_M}{3 m_N g_A}\left(\mathcal Q+2m_e -(E_{e1}+E_{e2})\frac{2 E_{e1}E_{e2}-m_e^2}{E_{e1}E_{e2}}\right)\right] \nonumber\\
 + & \frac{6}{g_A^2} \left(3 \epsilon_{GT} -\epsilon_F \right) \left( M_{GT}^K + M_{GT}^L \right) \bigg\} + \dots
\end{align}
Here $\mathcal Q =E_{11}+E_{22} - 2 m_e$ and the dots denote higher-order terms. We have inserted the Fermi function
\begin{eqnarray}\label{eq:fermi}
F(E,Z) &=& \left|\frac{2}{\Gamma(1+2 \gamma_0)}\right|^2 (2 p R)^{2(\gamma_0-1)} e^{\pi y} | \Gamma(\gamma_0+i y)|^2\,,
\end{eqnarray}
where $p = \sqrt{E^2 -m_e^2}$, $\gamma_0 = \sqrt{1-Z^2 \alpha^2}$, and $y = \alpha Z E/p$,
to account for the leading Coulomb corrections \cite{Simkovic:2018rdz}. This equation treats the nucleus as a point source, and it cures the divergence in the solution of the Dirac equation at $r=0$ by evaluating the wavefunction at a somewhat arbitrary ``nuclear radius'', $R$. Typically $R$ is chosen to be $R = 1.2 A^{1/3}$ fm \cite{Kotila:2012zza}, or, in the superallowed $\beta$ decay literature, $R = \sqrt{5/3}  \langle r^2_{\rm ch} \rangle^{1/2}$ \cite{Hayen:2017pwg,Wilkinson:1993hx}.

An EFT understanding of the Fermi function, its dependence on the nuclear scale $R$ and the interplay with nuclear-structure-dependent matrix elements were 
presented in Refs. \cite{Hill:2023acw,Hill:2023bfh,Cirigliano:2024msg}.
The Fermi function in Eq. \eqref{eq:fermi} can be refined by including the effects of the nuclear charge distribution and 
electron screening, using for example the Thomas-Fermi approximation where one introduces an ``effective'' charge $Z\to Z_{eff}\left(r,Z\right)$. We use Eq. \eqref{eq:fermi}
as our reference for the evaluation of phase space factors. 
Using the RADIAL Fortran package \cite{SALVAT2019165}, we checked that we can reproduce the $2\nu\beta\beta$ phase space factors in Ref. \cite{Kotila:2012zza,Simkovic:2018rdz}, if we adopt the prescription in \cite{Kotila:2012zza} and evaluate the electron wavefunctions at $r=R$.
While finite size and screening effects change the phase space by about 10\% \cite{Simkovic:2018rdz}, 
the effect on the normalized energy spectrum is about 0.1\% in the full energy range. At this level of accuracy, other missing corrections, such as radiative corrections, affect the spectrum. We therefore limit ourselves to the simple function in Eq. \eqref{eq:fermi}, but stress that future analyses should consider the theoretical error induced by missing electromagnetic corrections.

\section{The Lepton Energy Expansion }\label{sec:expansion}
For computing observables we wish to factor out the leptonic component from the NME. Following Ref.~\cite{Simkovic:2018rdz}, we expand Eq.~\eqref{eq:C2nu} with respect to $[\epsilon_{K,L}/\Tilde{A}]$, where $\tilde{A}=\left(E_n-\frac{1}{2}\left(E_i+E_f\right)\right)$ is the excitation energy with respect to the initial and final ground state.
\begin{eqnarray}
M^{K,L}_{GT} = \sum_{m=0}^{\infty} M^{(-2m-1)}_{GT}\left(\frac{\epsilon_{K,L}}{2m_e}\right)^{2m}\,,
\end{eqnarray}
where $M_{GT}^{(-2m-1)}$ contains nuclear energies and matrix elements
\begin{eqnarray}
 M_{GT}^{(-2m-1)} =m_e (2m_e)^{2m} \sum_n \frac{G_n}{\tilde{A}^{2m+1}}\,.
 \end{eqnarray}
In the next subsection we demonstrate that the lepton energy expansion converges rapidly. Expanding up to quartic terms we obtain
\begin{align}
& C_{2\nu} =      g_A^4 \left[M_{GT}^{(-1)} \right]^{2} \Delta_0
\Bigg( \mathcal A_0 +  \mathcal A_2 \, \xi_{31}  \frac{\Delta_2}{\Delta_0} +  \mathcal A_4 \left(\xi_{51} \frac{\Delta_2}{\Delta_0} + \frac{1}{3}\xi_{31}^2\right) + \mathcal A_{22}\frac{1}{3}\xi_{31}^2  + \frac{\mathcal A_M}{\Delta_0}\Bigg)+\dots
\end{align}
where the dots denote higher-order terms. We have defined here the functions of lepton energies
\begin{eqnarray}
\mathcal A_0 &=&1\,,\qquad 
\mathcal A_2 = \frac{\epsilon_K^2+\epsilon_L^2}{(2m_e)^2}\,,\qquad 
\mathcal A_4 = \frac{\epsilon_K^4+\epsilon_L^4}{(2m_e)^4}\,,\qquad
\mathcal A_{22}= \frac{\epsilon_K^2\epsilon_L^2}{(2m_e)^4} \nonumber\\
\mathcal A_M &=& \frac{2g_M}{3 m_N g_A} \frac{(E_{e1}+E_{e2})(2 E_{e1}E_{e2}-m_e^2)}{E_{e1}E_{e2}}\,,
\end{eqnarray}
and the combination of hadronic and nuclear matrix elements
\begin{eqnarray}\label{eq:Deltas}
\xi_{31} &=& \frac{{M_{GT}^{(-3)}}}{M_{GT}^{(-1)}}\,,\qquad \xi_{51} = \frac{{M_{GT}^{(-5)}}}{M_{GT}^{(-1)}}\,,\nonumber\\
\Delta_0 &=&  1 + \frac{4}{g_A^2}\frac{\left(3 \epsilon_{GT} -\epsilon_F \right)}{M_{GT}^{(-1)}} - \frac{2 g_M}{3 m_N g_A}(\mathcal{Q}+2 m_e)\,,\nonumber\\
\Delta_2 &=& 1 + \frac{2}{g_A^2}\frac{\left(3 \epsilon_{GT} -\epsilon_F \right)}{M_{GT}^{(-1)}}\,.
\end{eqnarray}
The WM and pionic corrections shift the overall decay rate through the factor $\Delta_0$ but isolating this term is difficult due to the poor theoretical knowledge of the effective combination $g_A^4 (M_{GT}^{(-1)})^2$. Furthermore, there are other terms at this order in the chiral expansion that modify the overall rate, such as, for example, two-body currents. 
The WM term will lead to corrections to the electron energy spectrum and we will discuss below how it can be isolated from the normalized differential decay rate. The pionic corrections are more difficult to isolate because the largest correction to the spectrum is degenerate with the ratio of NMEs $\xi_{31}$. While this ratio is significantly better known than the NMEs themselves, the uncertainty is still 
 quite large. At $\mathcal O(\epsilon_{K,L}^4)$ we see further pionic corrections proportional to $\Delta_2/\Delta_0$.

The total decay rate can be calculated from
\begin{eqnarray}\label{totdecay}
\frac{1}{\tau_{2\nu}} =\frac{\Gamma}{\ln 2}  =  g_A^4 \left(M_{GT}^{(-1)} \right)^2 \Delta_0 \left[G^{2\nu}_0+ \xi_{31} \frac{\Delta_2}{\Delta_0} G^{2\nu}_2 +G^{2\nu}_4 \left(\xi_{51} \frac{\Delta_2}{\Delta_0} + \frac{1}{3}\xi_{31}^2\right) + G^{2\nu}_{22}\frac{1}{3}\xi_{31}^2  +G^{2\nu}_M \right]\,,
\end{eqnarray}
in terms of the phase space integrals
\begin{eqnarray}
G^{2\nu}_{i} &=& \frac{1}{\ln 2} \frac{(G_F V_{ud})^4}{8 \pi^7 m_e^2}  \int^{E_i-E_f-m_e}_{m_e} d E_{e1} \int^{E_i-E_f-E_{e1}}_{m_e} d E_{e2} \int^{E_i-E_f-E_{e1}-E_{e2}}_{0} d E_{\nu1}\nonumber\\
&&\times E_{e1} p_{e1} E_{e2} p_{e2} E_{\nu 1}^2 E_{\nu 2}^2 \times F(E_{e1},Z_f) \times F(E_{e2},Z_f)\times \mathcal  A_{i}\,.
\end{eqnarray}
We often use a different set of kinematic variables and we define $\epsilon = E_{e1} + E_{e2} - 2 m_e$ and  $\Delta =(E_{e1}-E_{e2})/2$ as the total electron kinetic energy and electron energy difference.
We then write 
 \begin{eqnarray}
G^{2\nu}_{i} &=& \frac{1}{\ln 2}\frac{(G_F V_{ud})^4}{8 \pi^7 m_e^2}  \int_0^{\mathcal Q} d\epsilon \int^{\epsilon/2}_{-\epsilon/2} d \Delta \int^{\mathcal Q-\epsilon}_{0} d E_{\nu1}\nonumber\\
&&\times E_{e1} p_{e1} E_{e2} p_{e2} E_{\nu 1}^2 E_{\nu 2}^2 \times F(E_{e1},Z_f) \times F(E_{e2},Z_f)\times \mathcal  A_i\,.\label{Gamma}
\end{eqnarray}
The differential decay rate becomes
\begin{eqnarray}\label{totddecay}
\frac{d\Gamma}{d\epsilon} =  g_A^4 \left(M_{GT}^{(-1)} \right)^2 \Delta_0 \left[\frac{d G^{2\nu}_0}{d\epsilon}+ \frac{d G^{2\nu}_2}{d\epsilon}\xi_{31} \frac{\Delta_2}{\Delta_0} +\frac{d G^{2\nu}_4}{d\epsilon} \left(\xi_{51} \frac{\Delta_2}{\Delta_0} + \frac{1}{3}\xi_{31}^2\right) + \frac{d G^{2\nu}_{22}}{d\epsilon}\frac{1}{3}\xi_{31}^2  +\frac{d G^{2\nu}_M}{d\epsilon} \right]\,,
\end{eqnarray}
where
 \begin{eqnarray}
\frac{d G^{2\nu}_{i}}{d\epsilon} &=& \frac{1}{\ln 2}\frac{(G_F V_{ud})^4}{8 \pi^7 m_e^2}  \int^{\epsilon/2}_{-\epsilon/2} d \Delta \int^{\mathcal{Q}-\epsilon}_{0} d E_{\nu1}\nonumber\\
&&\times E_{e1} p_{e1} E_{e2} p_{e2} E_{\nu 1}^2 E_{\nu 2}^2 \times F(E_{e1},Z_f) \times F(E_{e2},Z_f)\times \mathcal A_i\,.\label{dGamma}
\end{eqnarray}

To factor out uncertainties in the overall NMEs, we consider the normalized differential decay rate (or the shape factor)
 \begin{eqnarray}
S(\epsilon)\equiv \frac{1}{\Gamma}\frac{d\Gamma}{d\epsilon}. 
\end{eqnarray}
In what follows, we will compute $S_i^{(n)}(\epsilon)$, where the superscript $\left(n\right)$ indicates the order of the $\epsilon^{2n}_{K,L}$ expansion, with $n \in \{0,1,2,\ldots\}$. A subscript is also added, $i \in \{0, WM, \pi, \chi\}$ which correspond to the uncorrected spectrum, the spectrum including only weak magnetism, only pion exchange, and both weak-magnetic and pionic corrections (chiral $\chi$). For example, Eq. \eqref{totddecay} after normalization by the total rate corresponds to $S^{\left(2\right)}_{\chi}$.

\subsection{Validation of the Expansion Method and the Role of Nuclear Uncertainties}
The lepton energy expansion method using QRPA matrix elements was introduced and discussed in detail in Ref. \cite{Simkovic:2018rdz}. We are interested in verifying the reliability of the expansion for the Nuclear Shell Model (NSM) and to assess the nuclear physics uncertainty on the $2\nu\beta\beta$ spectrum. 

Ref. \cite{Dekens:2024hlz} presented nuclear shell-model calculations of the matrix elements for the transitions 
from the $0^+$ ground states of  $^{76}\rm Ge$ and $^{76}$Se 
into $1^+$ states of the intermediate $^{76}$As, and, analogously, for the transitions of
$^{136}$Xe and $^{136}$Ba onto $^{136}$Cs. Details of the calculations, and the values of the reduced matrix elements used are given in Refs. \cite{Dekens:2024hlz,Jokiniemi:2022ayc}. 
Shell-model calculations overestimate the Gamow-Teller
matrix elements \cite{Menendez:2011qq}, and the effective Gamow-Teller matrix elements are obtained by multiplying the ``bare'' 
results in Ref. \cite{Dekens:2024hlz} by a quenching factor $q$.
The quenching factors have been estimated  as
$q_{^{76}\rm Ge} = 0.60$ and $q_{^{136}\rm Xe} = 0.45$ in Ref. \cite{Caurier:2011gi}, or, more conservatively, to be in the ranges 
$q_{^{76}\rm Ge}= \left[0.55,0.67\right]$ \cite{Jokiniemi:2022ayc},
$q_{^{136}\rm Xe}= \left[0.42,0.57\right]$ \cite{CoelloPerez:2018ghg}.
Accounting for the quenching factors and using the phase space factors from Ref.~\cite{Simkovic:2018rdz}, the resulting $2\nu\beta\beta$ half-lives are 
\begin{equation}\label{eq:total}
    T_{2\nu}(^{76}{\rm  Ge} ) = \left[1.5, 3.3\right] \cdot 10^{21} \; {\rm yr}\,, \qquad 
    T_{2\nu}(^{136}{\rm Xe}) = \left[0.6, 2.0\right] \cdot 10^{21} \; {\rm yr}\,, 
\end{equation}
compatible with the experimental results \cite{GERDA:2023wbr,EXO-200:2013xfn}
\begin{equation}
    T_{2\nu}(^{76}{\rm  Ge} ) = \left(2.022 \pm 0.042\right) \cdot 10^{21} \, {\rm yr}\,, \qquad 
    T_{2\nu}(^{136}{\rm Xe}) = \left(2.165 \pm 0.061  \right) \cdot 10^{21} \; {\rm yr}\,.
\end{equation}
The nuclear uncertainty on the total rate is a factor of two to three and thus sizable.

\begin{figure}
    \centering
    \includegraphics[width=0.49\linewidth]{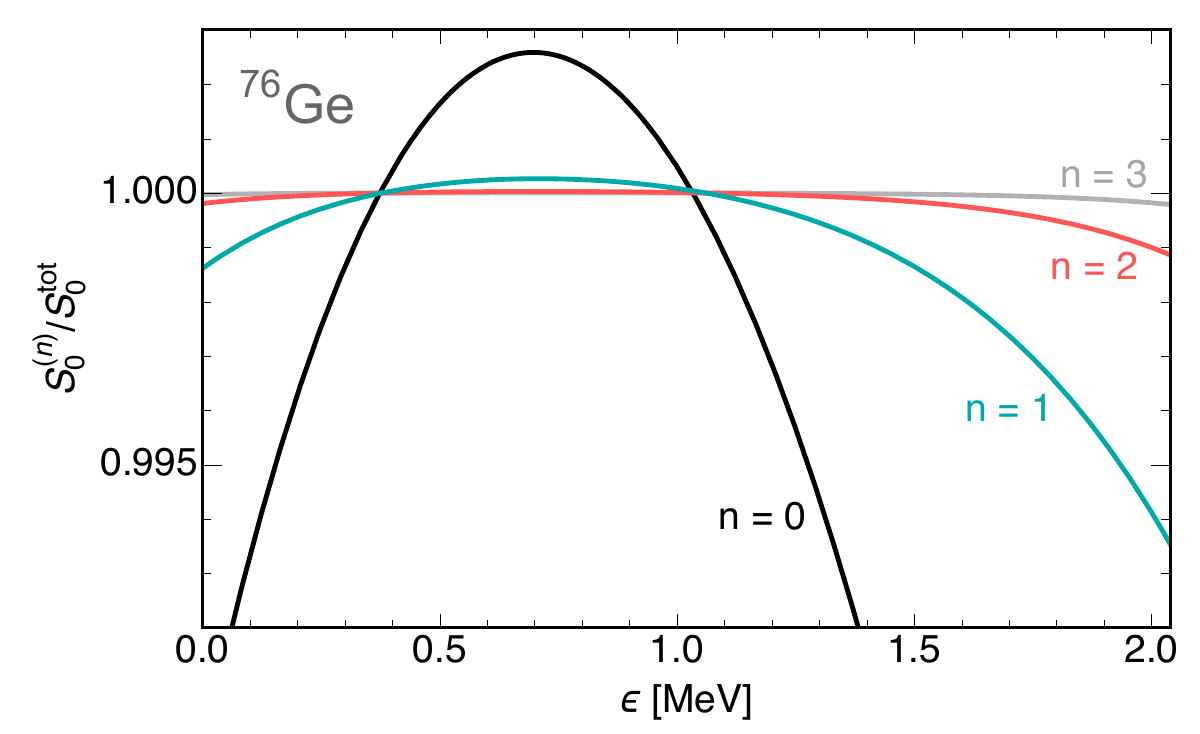}
    \includegraphics[width=0.49\linewidth]{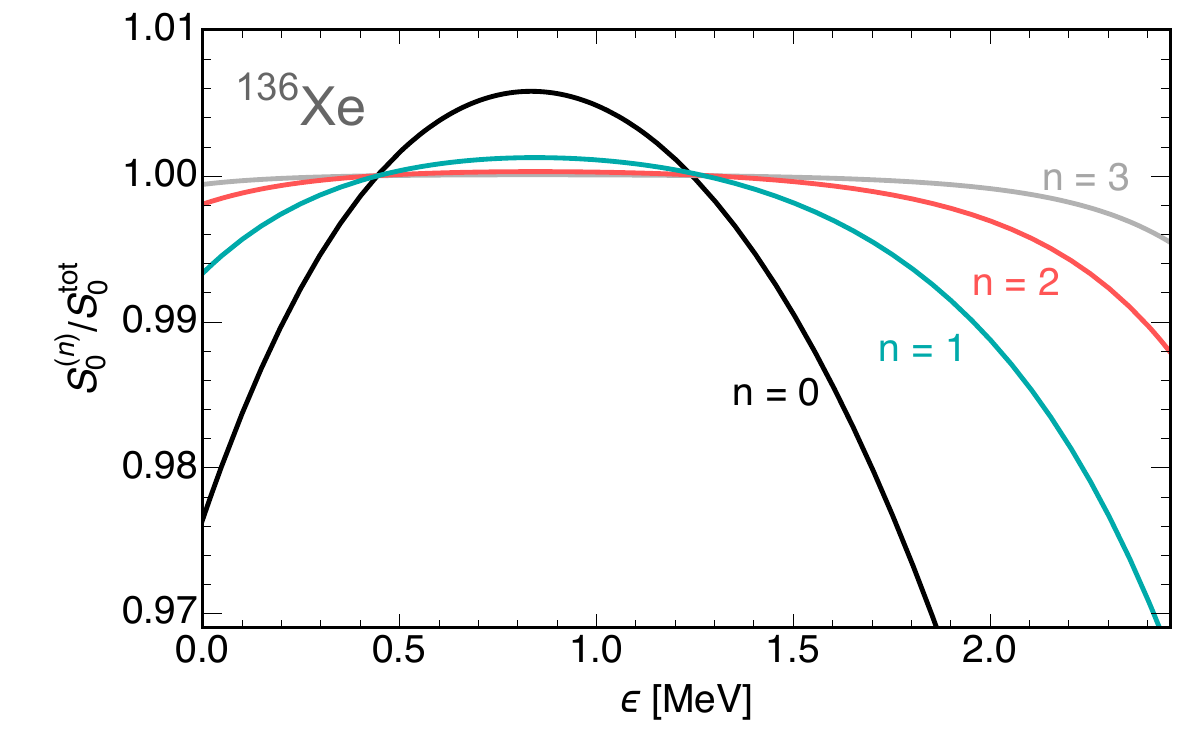}
    \caption{Convergence of the expansion for Ge (left) and Xe (right). We show the ratio of the expanded to the unexpanded shape factor, $S^{(n)}_0/S^{\mathrm{tot}}_0$, for order 0 (black), 1 (green), 2 (red), and 3 (gray). $S^{\mathrm{tot}_0}$ denotes the shape factor without expanding in lepton energies.}
    \label{fig:expansioncheck}
\end{figure}

To test the convergence of the epsilon expansion we first turn off the chiral and WM corrections and compare the unexpanded spectral shape $S(\epsilon)$ to $S^{(n)}(\epsilon)$ for $n=\{0,1,2,3,\ldots\}$ in Fig.~\ref{fig:expansioncheck}. We see convergence for both Ge and Xe, but to control the spectral shape at the per-mille level requires $n=2$ corrections for ${}^{76}$Ge and, in principle, $n=3$ corrections for ${}^{136}$Xe. Considering that the difference between $n=3$ and $n=2$ for ${}^{136}$Xe is only relevant near the end-point of the spectrum, we will expand up to $n=2$ in the rest of this work.

Each order in the lepton energy  expansion is associated with new ratios of NMEs such as $\xi_{31}$. The uncertainty on these ratios will make it harder to identify the chiral, WM or BSM corrections on the $2\nu\beta\beta$ spectrum. Within the QRPA method the uncertainty on the NME ratios can be estimated by comparing the calculated ratios for different amount of $g_A$ quenching \cite{Simkovic:2018rdz}. For each effective $g_A$ value, the amount of proton-neutron pairing in the Hamiltonian is adjusted to fit the $2\nu\beta\beta$ total rate, but this does not fix the values of the NME ratios. Using this variation, from the results presented in Ref. \cite{Simkovic:2018rdz} we obtain 
\begin{eqnarray}\label{xiQRPA}
{}^{76}\mathrm{Ge}(\mathrm{QRPA})&:&\qquad \xi_{31} = 0.11\pm 0.01\,,\qquad \xi_{51} = 0.021 \pm 0.004\,,\nonumber\\
{}^{136}\mathrm{Xe}(\mathrm{QRPA})&:&\qquad \xi_{31} = 0.32\pm 0.06\,,\qquad \xi_{51} = 0.10 \pm 0.02\,,
\end{eqnarray}
corresponding to $10\%$-$20\%$ nuclear uncertainty on the ratios.

\begin{figure}
    \centering
    \includegraphics[width=0.495\linewidth]{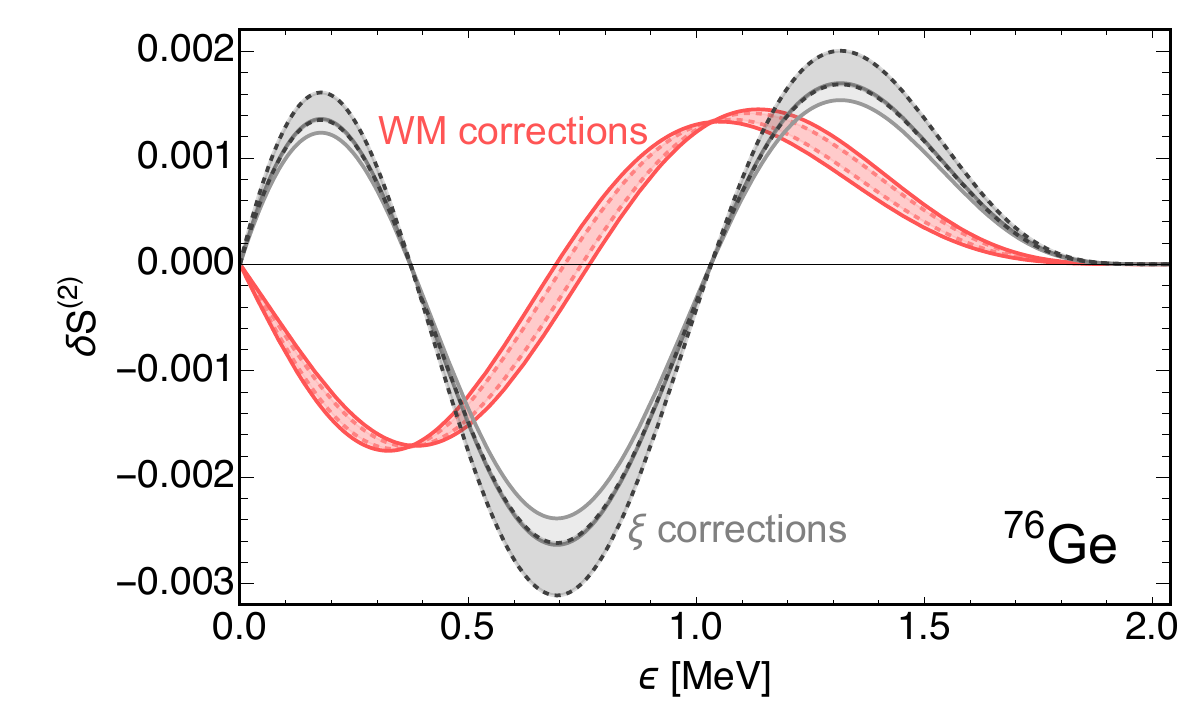}
    \hfill
    \includegraphics[width=0.495\linewidth]{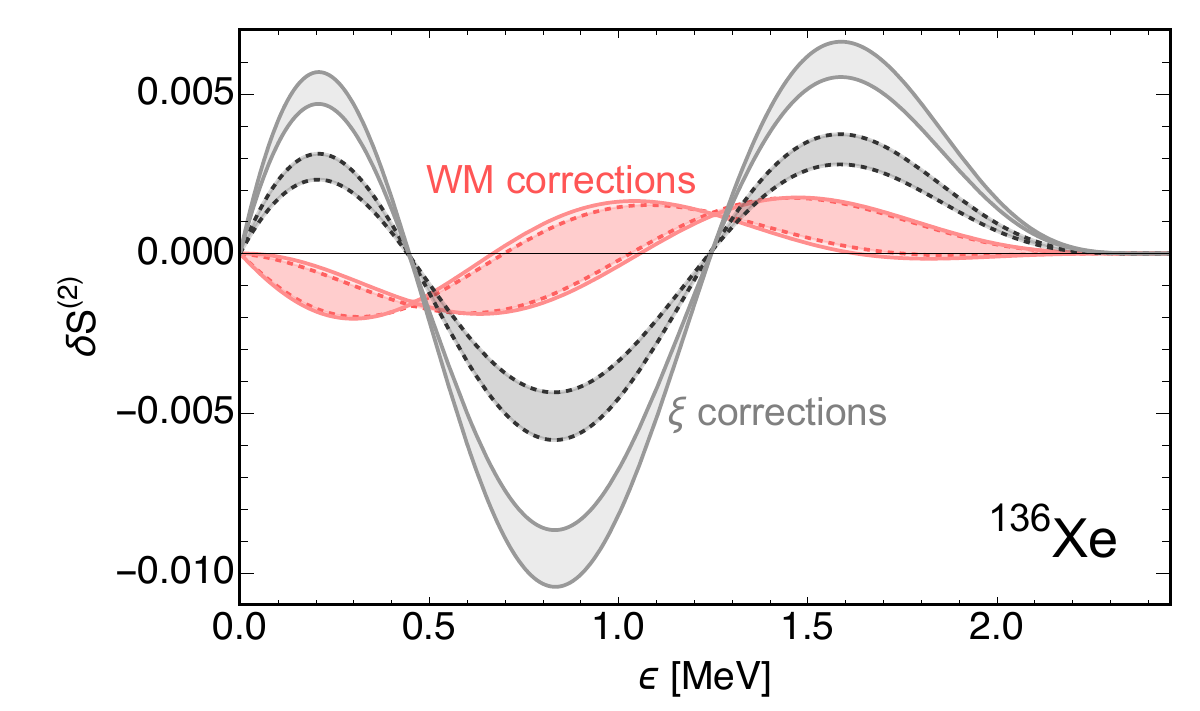}
    \caption{Effects of theoretical uncertainties on $\xi_{31}$ and $\xi_{51}$ on $\delta S^{(2)}$ for ${}^{76}$Ge (left) and ${}^{136}$Xe (right) in comparison with the effects of the correction induced by the weak magnetism term (red). We show results for both NSM (darker, dashed curves) and QRPA (lighter, solid curves).}
    \label{epsuncertain2}
\end{figure}

This is only the uncertainty within a given nuclear many-body method. We would like to compare this to NME ratios obtained within the nuclear shell model. Using the NMEs obtained in Ref.~\cite{Dekens:2024hlz} we can compute the central values given in Table~\ref{tab:comparison}. To assess the uncertainty we see how much the ratios vary if we change the individual transition matrix elements. We generate 5000 sets of matrix elements by multiplying 
each by a random number between 0.1 and 10, subjected to the restriction that the $2\nu\beta\beta$ half-life is within the range given in Eq. \eqref{eq:total}. We then use this set to compute an average $\xi_{ij}$. We obtain
\begin{eqnarray}\label{xiNSM}
{}^{76}\mathrm{Ge}(\mathrm{NSM})&:&\qquad \xi_{31} = 0.12\pm 0.02\,,\qquad \xi_{51} = 0.022 \pm 0.004\,,\nonumber\\
{}^{136}\mathrm{Xe}(\mathrm{NSM})&:&\qquad \xi_{31} = 0.16\pm 0.04\,,\qquad \xi_{51} = 0.042 \pm 0.012\,,
\end{eqnarray}
finding $20\%$-$30\%$ uncertainties. We stress that this is just a rough estimate of the theoretical uncertainty. 
While the ${}^{76}$Ge NME ratios are in good agreement with the QRPA results, the ${}^{136}$Xe ratios are smaller by about a factor 2. The mismatch for ${}^{136}$Xe indicates a disagreement between QRPA and the NSM regarding the importance of the low-energy excited states. An extraction of $\xi_{31}$ from results of the KamLAND-Zen experiment showed that part of the range of $\xi_{31}$ as predicted by the QRPA was partially excluded \cite{KamLAND-Zen:2019imh} but more data is necessary to draw a firmer conclusion. Below we will present results for the QRPA and NSM separately instead of attempting to combine the uncertainties from the two methods.
We focus here on QRPA and NSM NMEs, but the discussion can be applied to other many-body methods. In particular, it would be interesting to apply the same \textit{ab initio} methods used for $0\nu\beta\beta$ NMEs \cite{Belley:2023lec,Belley:2024zvt,Novario:2020dmr,Yao:2022usd} to $2\nu\beta\beta$, which might shed light on the issue of $g_A$ quenching, and provide a validation of the calculations of $\xi_{31}$ and $\xi_{51}$.

In Fig.~\ref{epsuncertain2} we show the resulting uncertainty on $\delta S^{(2)}_{0} \equiv (S^{(2)}_{0}- S^{(0)}_{0})$ due to the theoretical uncertainty on $\xi_{31}$ and $\xi_{51}$ for both QRPA and NSM. In particular, for ${}^{136}$Xe the discrepancies in the NMEs between the QRPA and the NSM are worrying. However, at certain specific energies, the uncertainties become small because the dependence on $\xi_{31}$ disappears. These nodes occur whenever 
\begin{equation}
\frac{d G_2^{2\nu}}{d\epsilon} \simeq  \frac{G_2^{2\nu}}{G_0^{2\nu}}\frac{d G_0^{2\nu}}{d\epsilon}\,,
\label{eq:node_equation}
\end{equation}
which depends just on phase space factors and not on nuclear theory. The nodes are interesting places to find deviations from the predicted $2\nu\beta\beta$ spectrum. We illustrate this by depicting the effect of WM corrections in red. The WM effects are comparable to the $\epsilon$ corrections and distinguishable despite nuclear uncertainties, in particular for energies around the nodes.

\section{A sensitivity study of the spectrum}\label{sec:sensitivity}

In this section, we discuss the quantitative impact of the chiral corrections on the (differential) decay rate. The total $2\nu\beta\beta$ rate is not precisely predicted by nuclear theory and suffers from a large theoretical uncertainty, as indicated in Eq.~\eqref{eq:total}. Nevertheless, it is interesting to study the impact of the chiral corrections on the total rate. To do so, we use the central value of $M_{GT}^{(-1)}$ and the $\xi$ ratios as given in Table~\ref{tab:comparison}. We give the resulting corrections to the lifetime in Table \ref{tab:gamma-table}. The chiral corrections on the lifetime are dominated by the pionic contributions and reduce the ${}^{76}$Ge lifetime by a $2\%$-$5$\% and of ${}^{136}$Xe by $7\%$-$13$\%. These shifts are sizable and indicate that it might be interesting to consider the remaining chiral corrections to the lifetime.

\begin{table}[t]
\centering
\begin{tabular}{|l|l|l|l|l|}
\hline
& $^{76}\rm Ge$ QRPA & $^{76}\rm Ge$ NSM & $^{136}\rm Xe$ QRPA  & $^{136}\rm Xe$ NSM \\
\hline
$T^{\left(2\right)}_0$ & $1.50$      & $2.13  $    & $1.79$        & $1.21$ \\
\hline
$T^{\left(2\right)}_\pi$ & $1.43$      & $2.08$    & $1.56$        & $1.12$ \\
\hline
$T^{\left(2\right)}_{\mathrm{WM}}$ & $1.51  $      & $2.14$    & $1.79$        & $1.21$ \\
\hline
$T^{\left(2\right)}_\chi$ &$ 1.43     $  & $2.08$     & $1.56$     & $1.12$\\
\hline
\end{tabular}
\caption{Decay rates in units of $10^{21}$ yr with ($T^{\left(2\right)}_{i\neq 0}$) chiral and without ($T^{\left(2\right)}_0$) chiral corrections using Eqs. \eqref{xiQRPA} and \eqref{xiNSM}. We have used  $M_{GT}^{(-1)}$ and $\xi$ values given in Table~\ref{tab:comparison}. }
\label{tab:gamma-table}
\end{table}

\begin{figure}[t]
    \centering
    \includegraphics[width=0.49\linewidth]{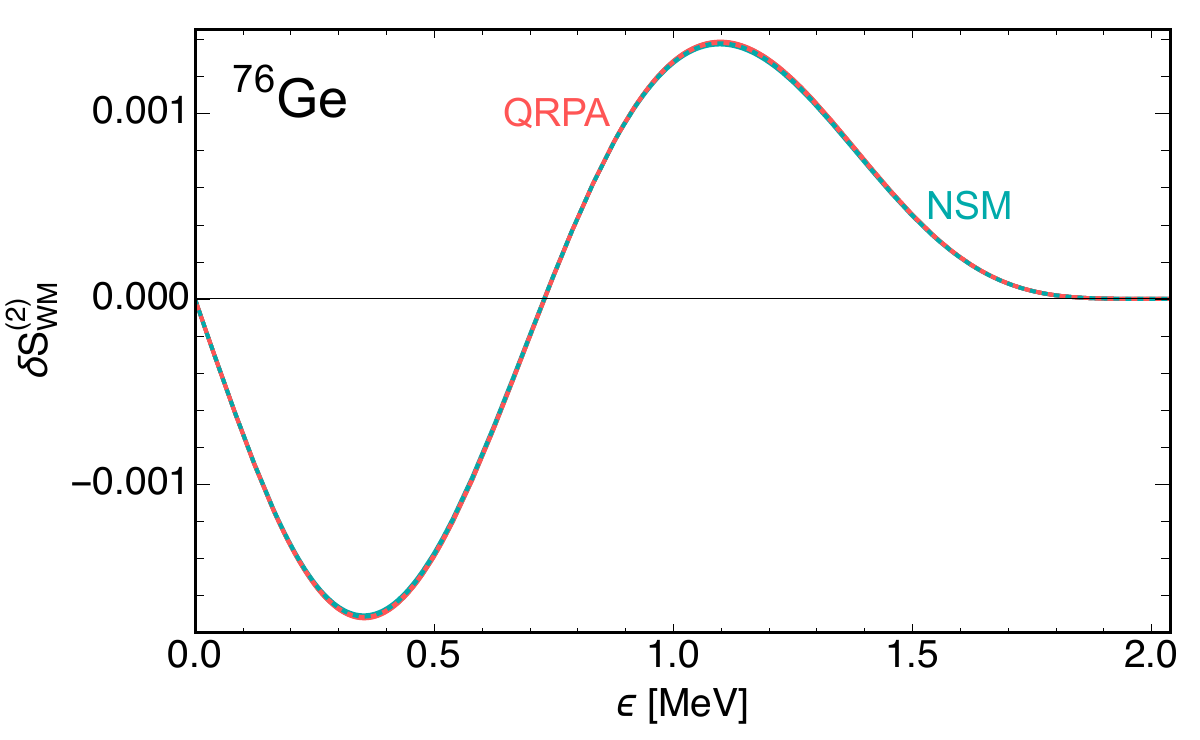} \hfill
    \includegraphics[width=0.49\linewidth]{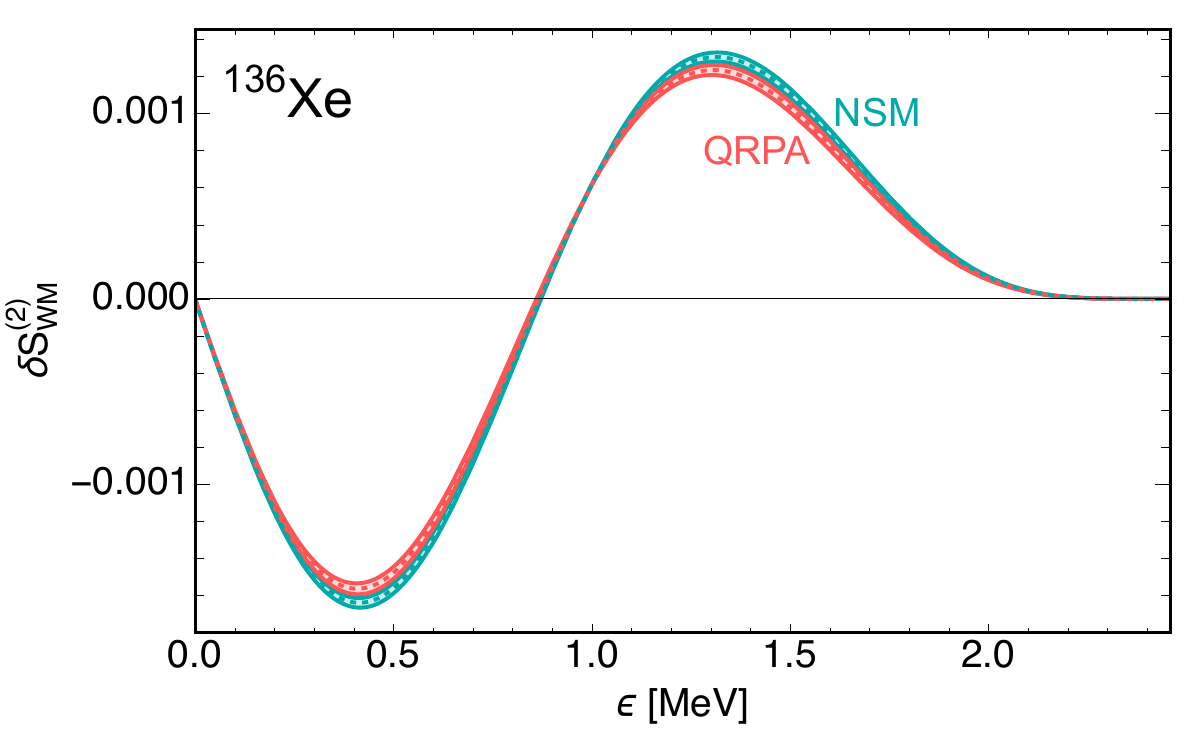}
    \includegraphics[width=0.49\linewidth]{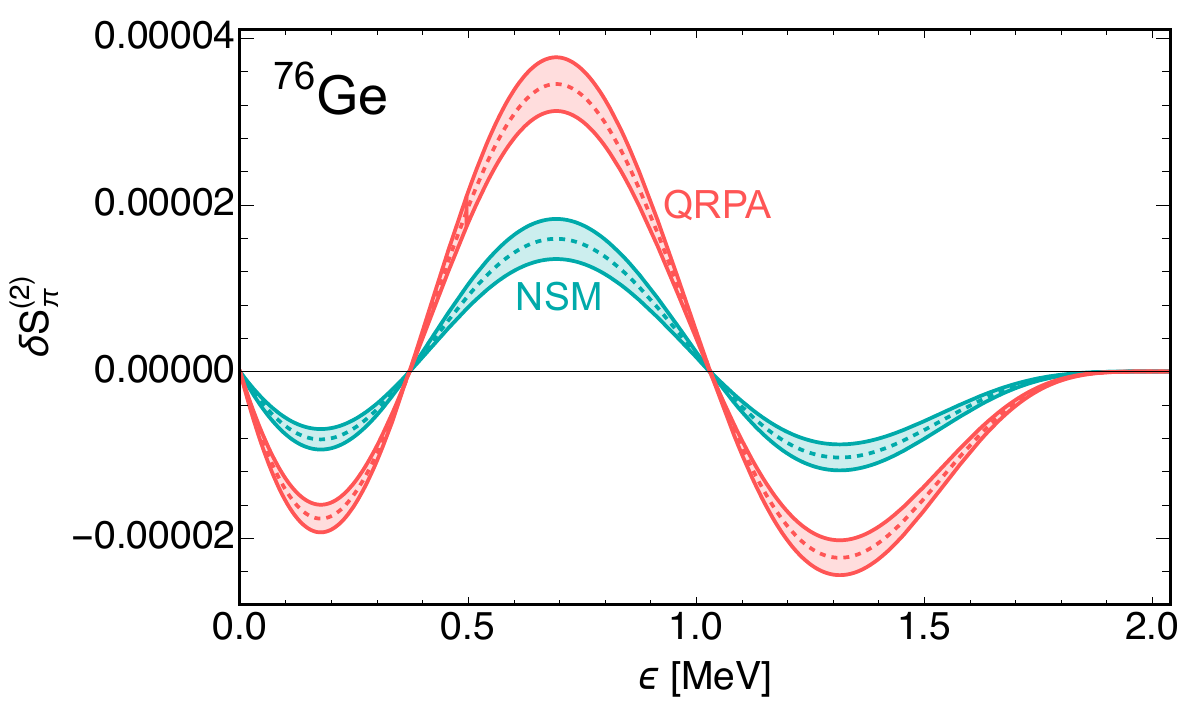} \hfill
    \includegraphics[width=0.49\linewidth]{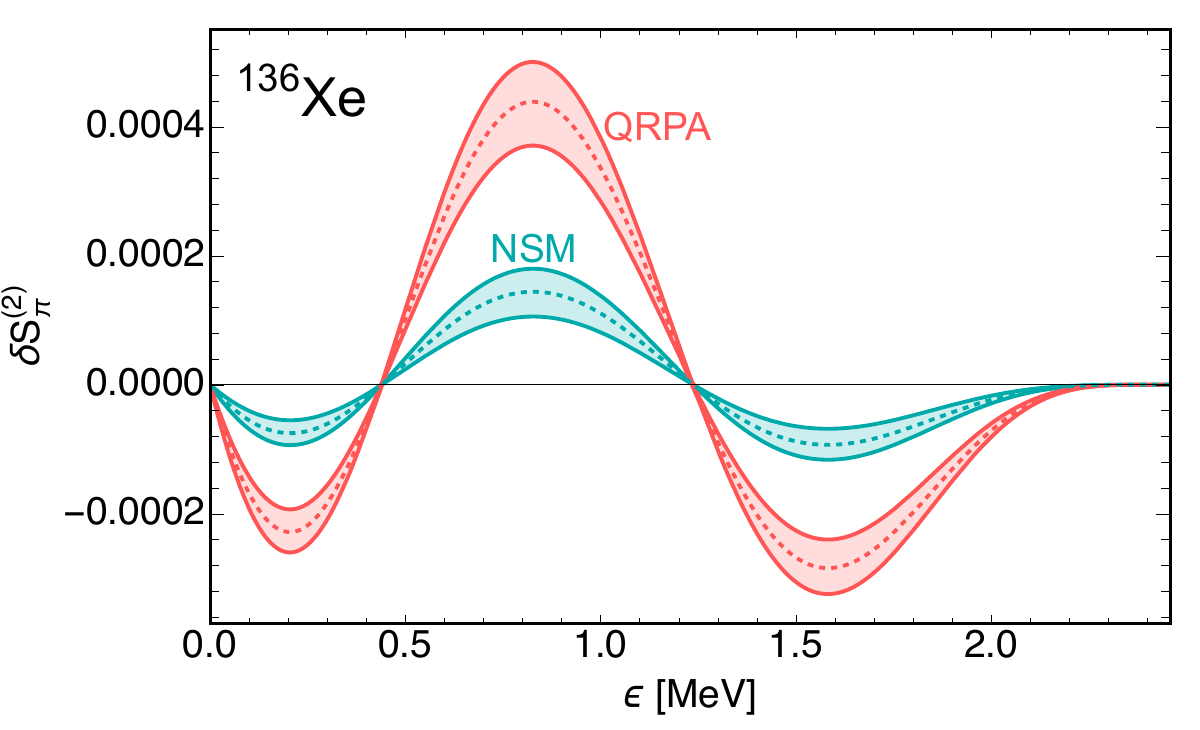}
    \caption{Distortion in the spectral shape due to weak magnetism term (upper panel) and pion corrections (lower panel) for Germanium (left) and Xenon (right) with the $\xi_{31}$ and $\xi_{51}$ parameters varied within the ranges given in Eqs.~\eqref{xiNSM} (NSM, depicted in blue) and in Eqs.~\eqref{xiQRPA} (QRPA, depicted in red).}
    \label{fig:WMold}
\end{figure}

Corrections to the total rate are hard to isolate due to the large nuclear uncertainty. To quantify deviations on the shape factor we consider both the difference of spectra and relative fraction:
\begin{eqnarray}\label{deltaSdef}
\delta S_{i}^{(m)}(\epsilon) = S_{i}^{(m)}(\epsilon) - S_{0}^{(m)}(\epsilon)\,,\quad \delta \bar S_{i}^{(m)}(\epsilon) = \delta S_{i}^{(m)}(\epsilon)/S_{0}^{(m)}(\epsilon)\,.
\end{eqnarray}
Although easy to compute with the expressions given above, the interpretation in terms of measurable quantities requires careful thinking regarding nuclear uncertainties.
We depict the distortion of the energy electron spectrum in Fig.~\ref{fig:WMold}. The width of the bands is obtained by varying $\xi_{31}$ and $\xi_{51}$ by their ranges specified in Eqs.~\eqref{xiQRPA} and $\eqref{xiNSM}$. It is important to stress that while we vary $\xi$'s in this range we use the same value in both $S_{0}^{(m)}$ and $S_{i}^{(m)}$. That is, the uncertainty is correlated leading to rather thin bands in Fig.~\ref{fig:WMold}. The effect of WM on the $^{76}$Ge and $^{136}$Xe spectra is shown in the upper panel of Fig. \ref{fig:WMold}.
WM shifts the spectrum's peak to higher energies and causes few per-mille changes to the shape of the spectrum. The pion corrections, shown in the lower panel,  modulate the shape, with enhancement at the center, attenuation at roughly two- and four-fifths of the spectrum, and nodes at the endpoints and in between the attenuation and enhancement. For ${}^{76}$Ge the pionic corrections are very small, but they are comparable to WM for ${}^{136}$Xe. However, the pionic corrections should only be seen as rough estimates as the short-distance contributions are not included and could change the overall size of the effect. 

\begin{figure}
    \centering
    \includegraphics[width=0.49\linewidth]{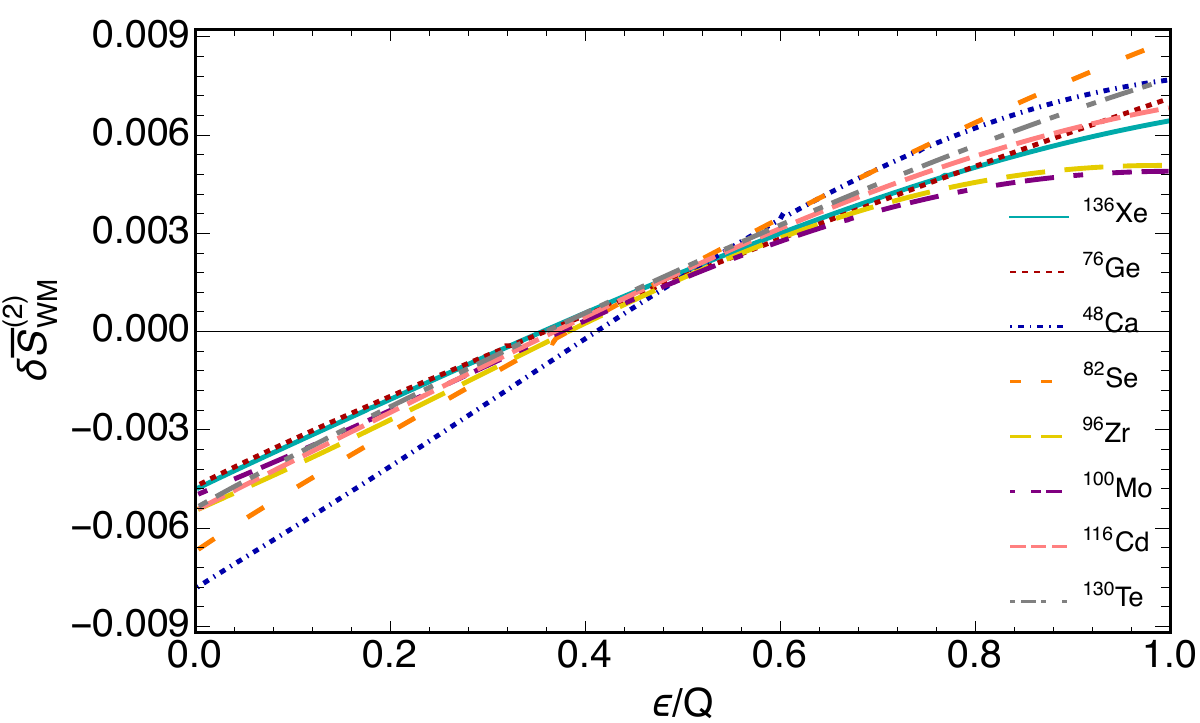} \hfill
    \includegraphics[width=0.49\linewidth]{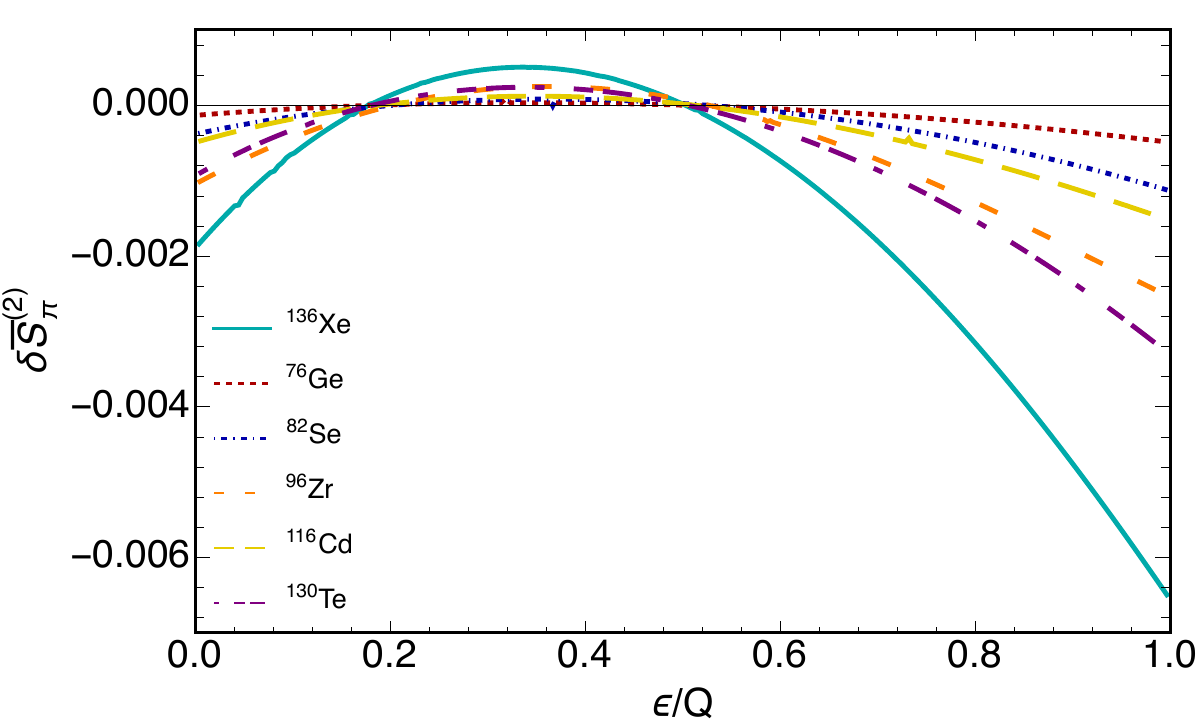}
    \caption{The weak-magnetic (left) and pionic (right) corrections to the normalized spectral shape for various isotopes.}
    \label{fig:multi_iso}
\end{figure}

While we focus on $^{76}$Ge and $^{136}$Xe in this paper for brevity, the identified spectral distortions also appear for other isotopes that undergo $2\nu\beta\beta$. We show the WM and pionic corrections in the left and right panels of Fig.~\ref{fig:multi_iso}, respectively. We have used $\xi$ values from Ref.~\cite{Simkovic:2018rdz} and the $0\nu\beta\beta$ NMEs necessary to evaluate $\epsilon_{F,GT}$ from Ref.~\cite{Cirigliano:2018yza},
which used the QRPA calculation of Ref. \cite{Hyvarinen:2015bda}. 
We have not included nuclear uncertainty bands to avoid cluttering the plot. The WM effects are fairly isotope independent, ranging from $-0.7\%$ to $+0.7\%$ over the energy range. The pionic corrections show a smaller variance for ${}^{76}$ Ge and ${}^{82}$ Se but a larger variance for heavier isotopes ${}^{130}$Te and ${}^{136}$Xe (up to $-0.8\%$ near the end point of the spectrum).

While this shape modulation is non-trivial, the uncertainty in the value of $\xi_{31}$ and $\xi_{51}$ complicates things. The extrema of the pionic corrections are washed out by the uncertainty in $\xi_{31}$, which are convolved with the pionic corrections via the product $\xi_{31} \Delta_2/\Delta_0$. To identify WM, chiral, or BSM corrections from accurate measurements of the spectrum, we either have to subtract $S_0(\epsilon)$ from the data and compare the resulting difference $\delta S_{\rm exp} = S_{\rm exp} - S_0$ with $\delta S_i$, or perform a joint fit to $\xi_{31}$, $\xi_{51}$, and the parameters in chiral corrections. The former procedure would add a theoretical error to $\delta S_{\rm exp}$. To model this theory error, and to assess whether the chiral corrections can be distinguished from 
the $\xi$ corrections to the spectrum, we treat the $\xi$s as independent normally distributed variables\footnote{We assume that the $\xi$s are uniformly distributed in the ranges specified by Eqs.~\eqref{xiQRPA} and \eqref{xiNSM}. We then compute the associated standard deviation for a normal distribution by taking the uniform spread divided by $\sqrt{12}$ to align the variance of the normal and uniform distribution. For example, for ${}^{76}$Ge and the QRPA method we obtain $\sigma (\xi_{31})=0.02/\sqrt{12}=0.0058$.} in $S_i(\epsilon)$ and $S_0(\epsilon)$, and then compute the resulting uncertainty on $\delta S_i(\epsilon)$. This procedure is equivalent to considering $\delta S_i - \delta S_{\rm exp}$, with the assumption that experimental data have no error, and their central value exactly agrees with $S_0(\epsilon)$, or $\delta S_{\rm exp} = 0$.
If $\delta S_i(\epsilon)$ calculated in this way is incompatible with zero, it would indicate that the effect is not overwhelmed by the uncertainties on $\xi$, and thus observable, provided that the experimental measurements have comparable accuracy.

\begin{figure}[t]
    \centering
    \includegraphics[width=0.49\linewidth]{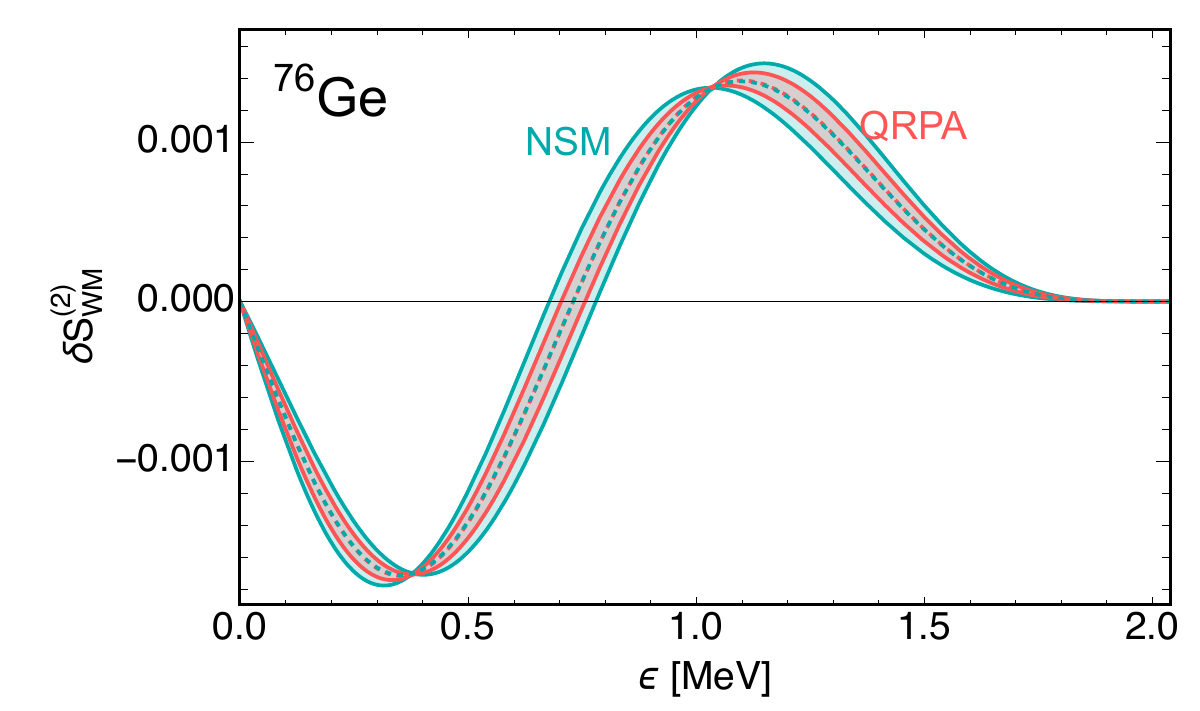}
        \includegraphics[width=0.49\linewidth]{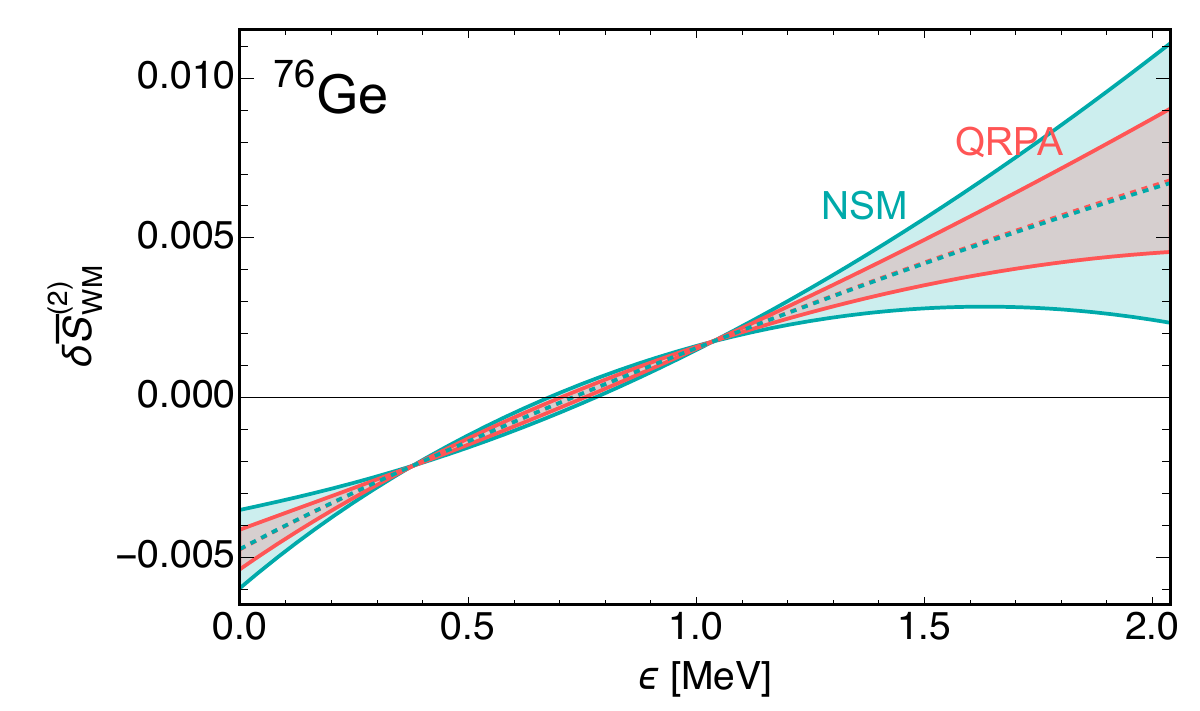}
        \includegraphics[width=0.49\linewidth]{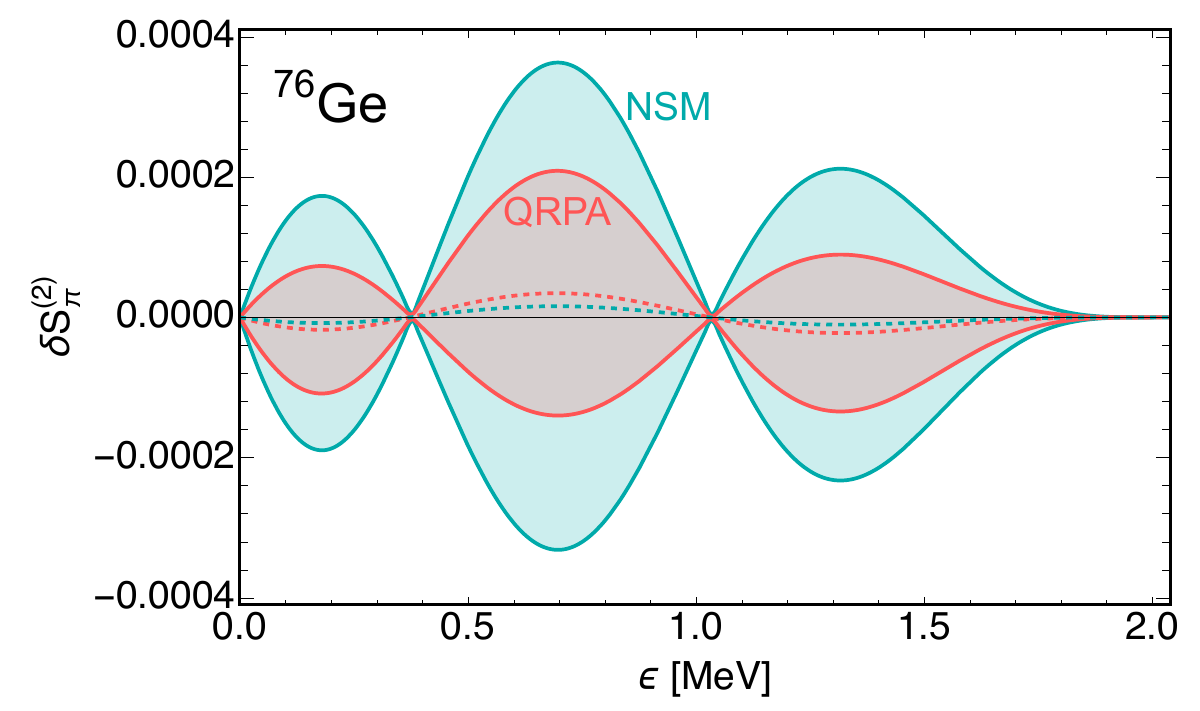}
        \includegraphics[width=0.49\linewidth]{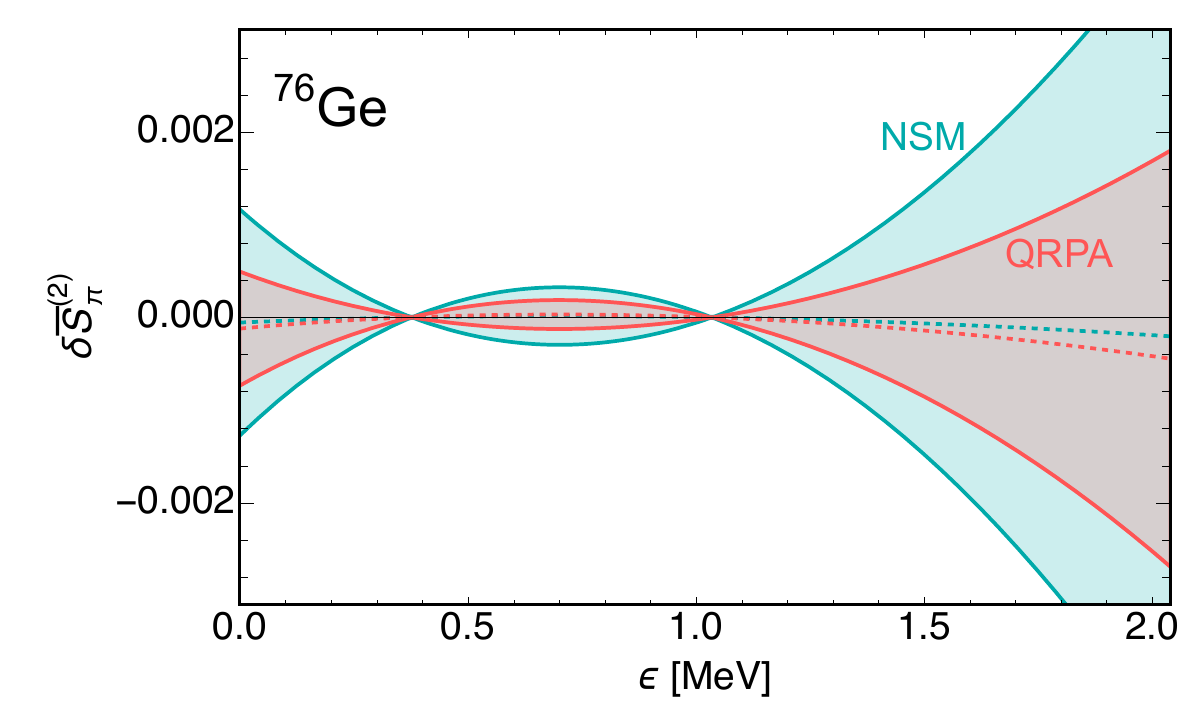}
    \caption{ Top: Absolute (left) and normalized (right) shift in the spectral shape due to WM (top) and pion (bottom) corrections for ${}^{76}$Ge for QRPA (red) and NSM (blue) NMEs. The procedure to obtain the uncertainty bands is described in the main text.}
    \label{fig:WMnew}
\end{figure}

\begin{figure}[t]
    \centering
    \includegraphics[width=0.49\linewidth]{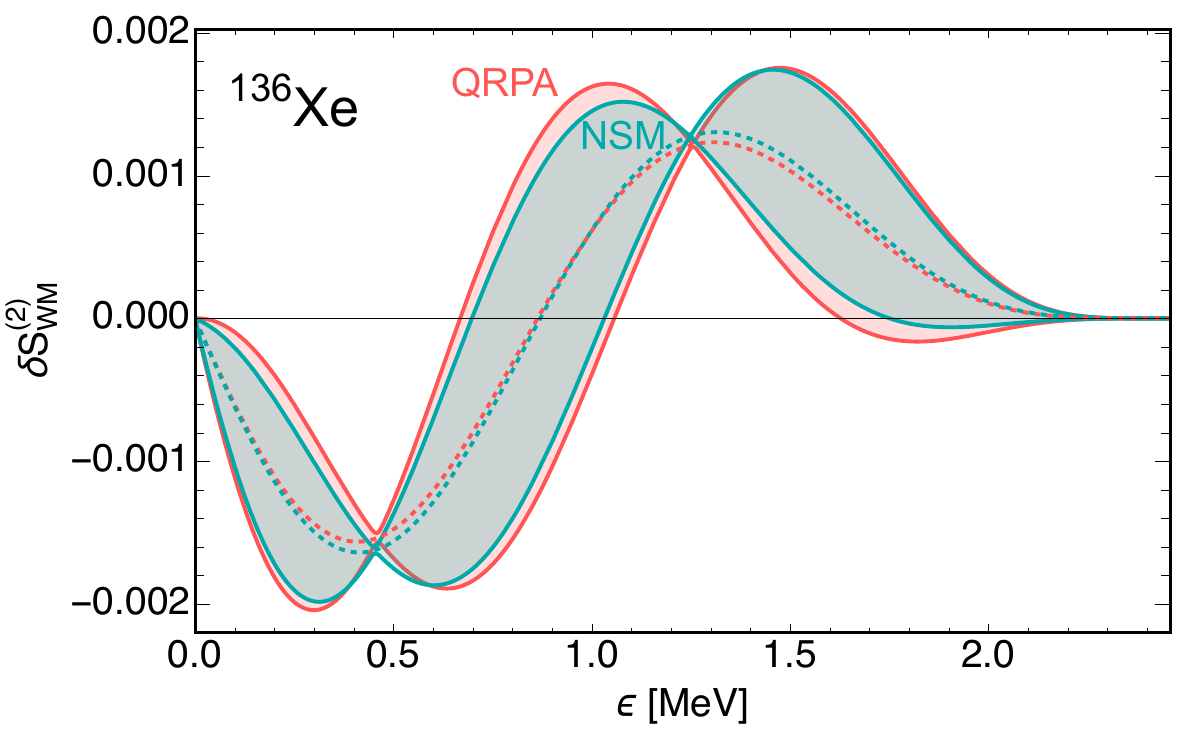}
        \includegraphics[width=0.49\linewidth]{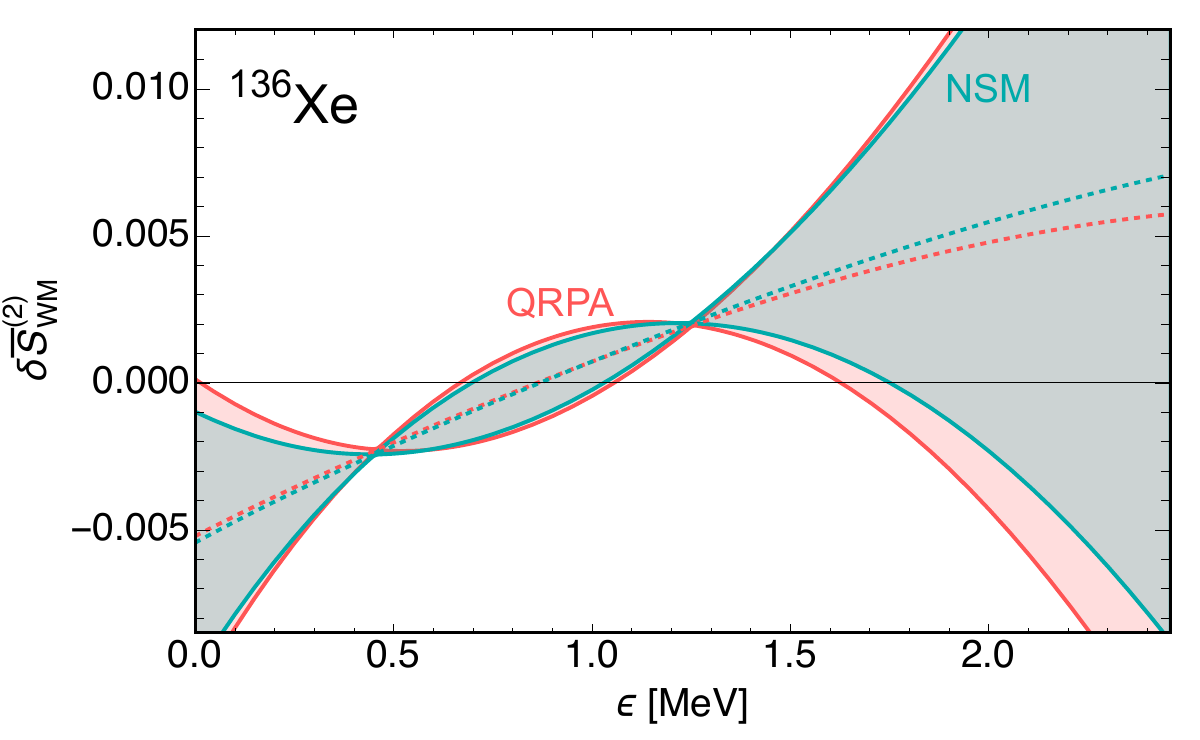}
        \includegraphics[width=0.49\linewidth]{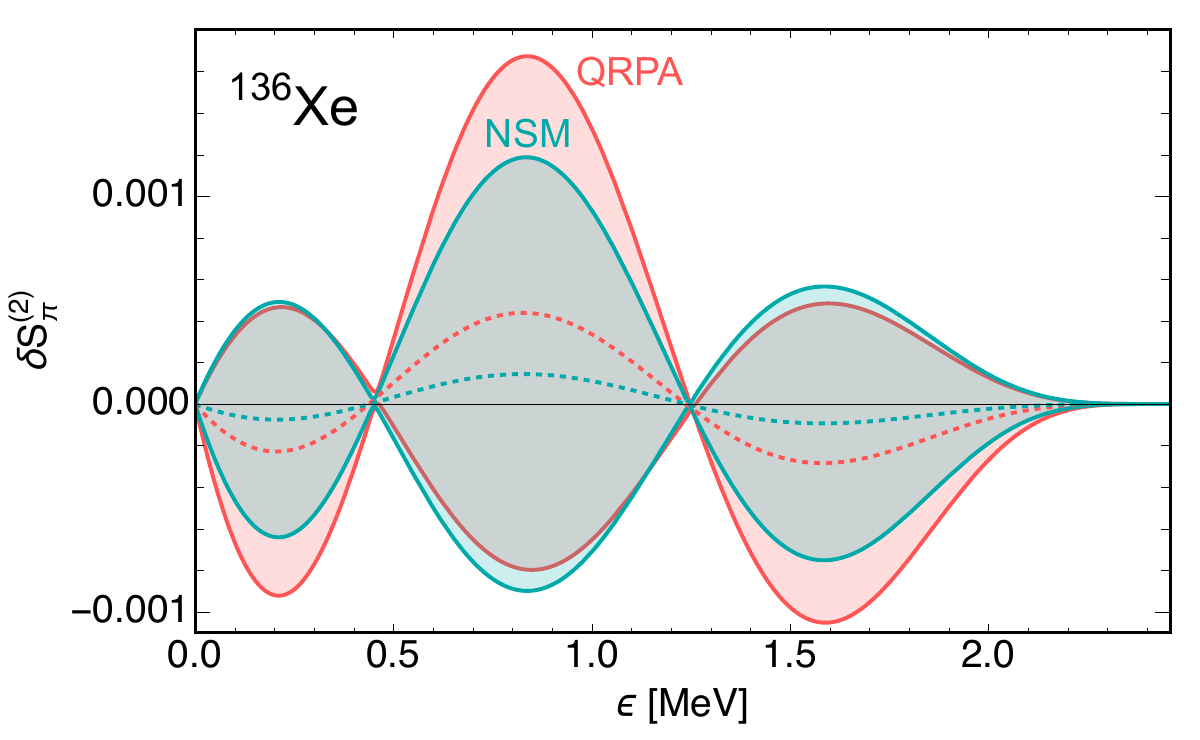}
        \includegraphics[width=0.49\linewidth]{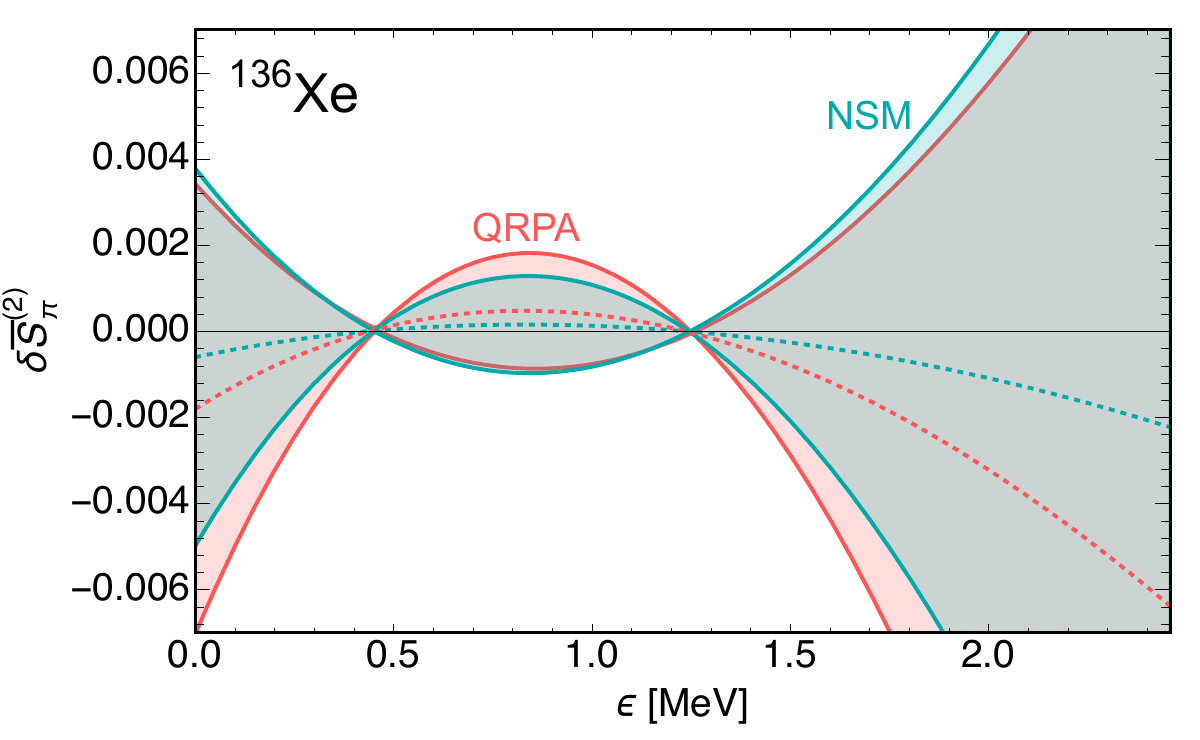}
    \caption{Top: Absolute (left) and normalized (right) shift in the spectral shape due to WM (top) and pion (bottom) corrections for ${}^{136}$Xe for QRPA (red) and NSM (blue) NMEs. The procedure to obtain the uncertainty bands is described in the main text.}
    \label{fig:WMnewXe}
\end{figure}

The resulting nuclear uncertainty is shown by the bands in Figs.~\ref{fig:WMnew}, for $^{76}$Ge, and \ref{fig:WMnewXe} for $^{136}$Xe. For both isotopes, the WM effects are still distinguishable, in particular for electron energies around the nodes identified in Eq. \eqref{eq:node_equation}. The pionic corrections are washed out with current $\xi_{31}, \xi_{51}$ uncertainties and are unlikely to be measured without greater control of, in particular, $\xi_{31}$.

%

\section{Beyond-the-Standard-Model Contributions to \vvbb}\label{sec:Beyond-the-Standard-Model Contributions to 2nu}
Considering the large amount of experimental activity in the field of $2\nu\beta\beta$, it would be interesting to use the measured electron spectra as a test of the SM. This has already been done to look for sterile neutrinos, majorons, and 
CPT violation \cite{Blum:2018ljv,Deppisch:2020mxv,Agostini:2020cpz,Bolton:2020ncv,Agostini:2015nwa,CUPID:2024qnd,GERDA:2022ffe,Kharusi:2021jez,CUPID-0:2022yws,Diaz:2013ywa,Diaz:2013saa}.
In this section, we review how various BSM effects could modify the electron spectrum and how this compares to the chiral corrections identified above. 

\subsection{\vvbb with Sterile Neutrinos}
One interesting possibility is that in $2\nu\beta\beta$ a so-called sterile neutrino is produced with a mass below the $\mathcal Q$-value of the reaction. In this case, the total measured differential decay rate is given by 
\begin{align}
\label{eq:total_dist}
	\frac{d\Gamma^{2\nu}(m_N, |V_{eN}|)}{d \epsilon} 
	= (1 - |V_{eN}|^2)^2 \frac{d\Gamma}{d \epsilon} 
	+ 2(1-|V_{eN}|^2)|V_{eN}|^2 \,\frac{d\Gamma^{\nu N}(m_N)}{d\epsilon}\,,
\end{align}
in terms of the mass of the sterile neutrino $m_N$ and the mixing angle $V_{eN}$. 
The rate $\Gamma^{\nu N}$ is defined for the process where a light and a sterile neutrino is produced. It can be calculated from 
\begin{eqnarray}
\frac{\Gamma^{\nu N}}{\ln 2}  =  g_A^4 \left(M_{GT}^{(-1)} \right)^2 \left[G^{\nu N}_0+ \xi_{31} G^{\nu N}_2 +G^{\nu N}_4 \left(\xi_{51} + \frac{1}{3}\xi_{31}^2\right) + G^{\nu N}_{22}\frac{1}{3}\xi_{31}^2 \right]\,,
\end{eqnarray}
where the phase space integrals are now defined as
 \begin{eqnarray}
G^{\nu N}_{i} &=& \frac{1}{\ln 2}\frac{(G_F V_{ud})^4}{8 \pi^7 m_e^2}  \int_0^{\mathcal Q-m_N} d\epsilon \int^{\epsilon/2}_{-\epsilon/2} d \Delta \int^{\mathcal Q-\epsilon}_{m_N} d E_{N}\nonumber\\
&&\times E_{e1} p_{e1} E_{e2} p_{e2} E_{\nu 1}^2 E_{N} \sqrt{E_N^2-m_N^2} \times F(E_{e1},Z_f) \times F(E_{e2},Z_f)\times A_i\,,\label{Gamma2}
\end{eqnarray}
where $E_{\nu 1} = (E_i - E_f  - E_{e1} - E_{e2} -E_N)$ is implied.   We define the shape factor as
\begin{eqnarray}
S_N(\epsilon)= \frac{1}{\Gamma^{2\nu}(m_N, |V_{eN}|)}	\frac{d\Gamma^{2\nu}(m_N, |V_{eN}|)}{d \epsilon} \,.
\end{eqnarray}

\begin{figure}[t]
    \centering
    \includegraphics[width=0.49\linewidth]{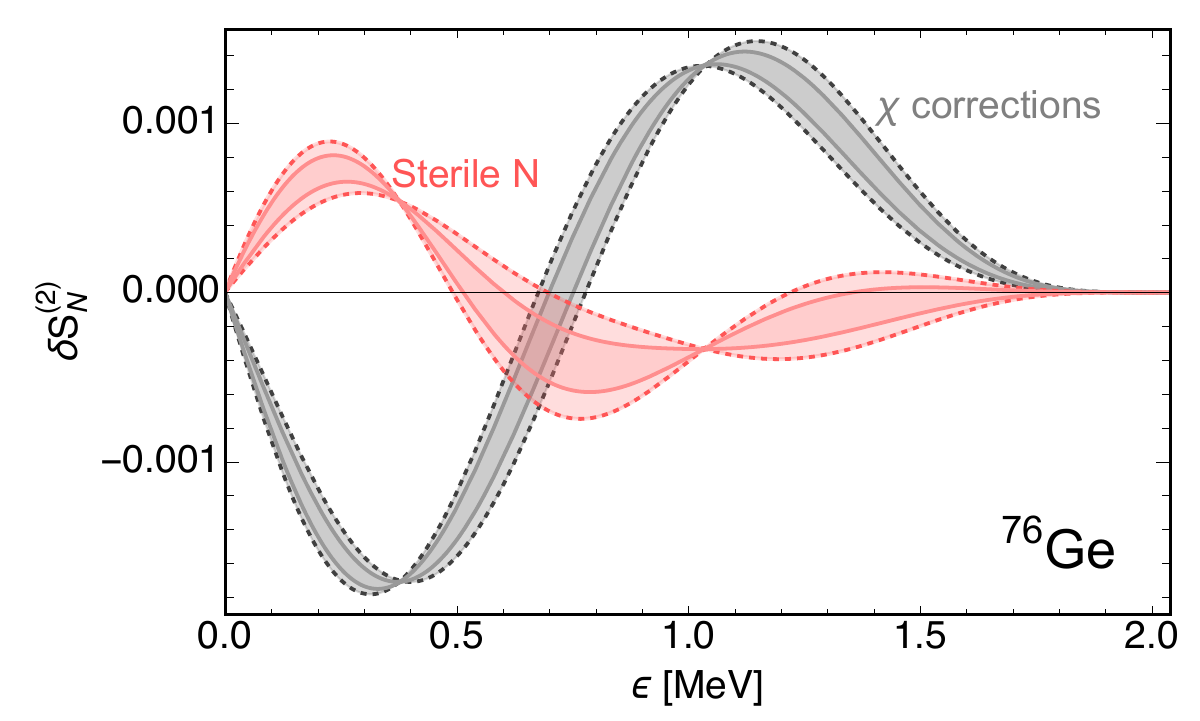}
    \includegraphics[width=0.49\linewidth]{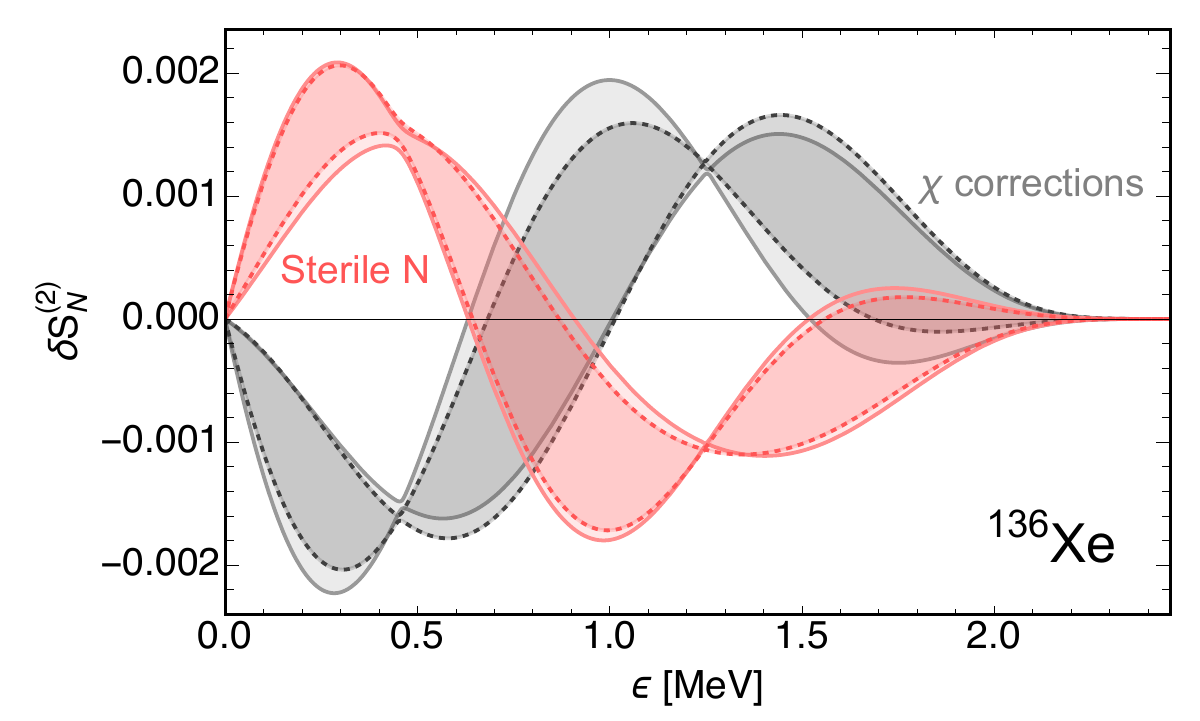}
    \caption{Shape-factor deviations $\delta S_N$ stemming from the chiral corrections in comparison with deviations introduced by a hypothetical sterile-neutrino contribution with mass $M_N=1$ MeV and active-sterile mixing $|V_{eN}|^2=0.1$ for ${}^{76}$Ge (left) and ${}^{136}$Xe (right).}
    \label{fig:sterile}
\end{figure}

As an example we depict 
\begin{eqnarray}
\delta S_{N}(\epsilon) = S_N(\epsilon) - S_0(\epsilon)\,,
\end{eqnarray}
for a sterile neutrino with mass $m_N=1$ MeV and a mixing angle $|V_{eN}|^2=0.1$ in Fig.~\ref{fig:sterile}. This value of mixing angle roughly corresponds to limits set by existing $2\nu\beta\beta$ analyses \cite{GERDA:2022ffe,CUPID:2024qnd}, although we stress that stronger limits are presently set by single $\beta$-decay spectrum analyses \cite{Deutsch:1990ut,Bolton:2019pcu}. The spectrum distortion from a sterile neutrino with this mass and mixing is of similar size as the chiral corrections identified in this work. 
The deviations from the two sources (sterile neutrino and chiral) pull the peak of the spectrum in opposite directions and, for certain parameter setting, could result in mutual cancellation. To be able to resolve sterile neutrino effects in the \vvbb spectrum or to set more stringent limits, the chiral corrections must be included in the analysis.

\subsection{Non-Standard Charged-Current Interactions}

Another possible form of BSM physics arises from non-standard charged-current interactions. Several options are possible \cite{Lee:1956qn,Cirigliano:2012ab}, but we focus here on tensor interactions of left-handed neutrinos and axial and vector couplings of massless right-handed neutrinos. 
In the conventions of Ref. \cite{Cirigliano:2012ab}, these interactions are captured by the Lagrangian
\begin{equation}
    \mathcal L = - \frac{4 G_F}{\sqrt{2}} V_{ud} \Big\{ 
    \epsilon_T \bar e_R \sigma^{\mu \nu}  \nu_L \, \bar u_R \sigma_{\mu \nu} d_L + \tilde \epsilon_R  \bar e_R \gamma^\mu  \nu_R \, \bar u_R \gamma_\mu d_R + 
    \tilde \epsilon_L  \bar e_R \gamma^\mu  \nu_R \, \bar u_L \gamma_\mu d_L
    \Big\}.
    \label{eq:bsmlagrangian}
\end{equation}
The couplings $\tilde\epsilon_R$ and $\tilde\epsilon_L$ are related to
$\epsilon_{RR}$ and
$\epsilon_{LR}$
introduced in Ref. \cite{Deppisch:2020mxv}
by
$\epsilon_{LR} = -\tilde{\epsilon}_L$,
$\epsilon_{RR} = -\tilde{ \epsilon}_R$.
The operators in Eq. \eqref{eq:bsmlagrangian} can be mapped onto gauge invariant operators in the SM Effective Field Theory (SMEFT) \cite{Grzadkowski:2010es} or 
sterile neutrino extended SMEFT
($\nu$SMEFT) \cite{delAguila:2008ir}. In particular, $\epsilon_T$
and $\tilde{\epsilon}_R$ correspond, respectively, to semileptonic four-fermion operators in SMEFT and $\nu$SMEFT, while $\tilde{\epsilon}_L$ emerges from a coupling of the $W$ boson to right-handed neutrinos and electrons. 
These interactions can be constrained in low-energy $\beta$ decays \cite{Cirigliano:2012ab,Cirigliano:2013xha,Falkowski:2020pma,Dekens:2020ttz} and $0\nu\beta\beta$ decays \cite{Dekens:2021qch}. $\epsilon_T$ and $\tilde{\epsilon}_R$ can also be constrained by neutral and charged-current Drell-Yan at the LHC
\cite{Cirigliano:2012ab,Alioli:2018ljm}, while 
$\tilde{\epsilon}_L$ can be indirectly constrained by its contribution to lepton dipole moments \cite{Cirigliano:2021peb}.
Considering only low-energy observables, the global analysis of Ref. \cite{Falkowski:2020pma}  yields, at the 90\% CL
\begin{equation}\label{eq:bounds}
 - 1.2 \cdot 10^{-3}     <  \epsilon_T  < 1.1 \cdot 10^{-3}, \qquad 
 |\tilde{\epsilon}_L + \tilde{\epsilon}_R | < 0.09, \qquad |\tilde{\epsilon}_L -\tilde{\epsilon}_R | < 0.08.
\end{equation}

The corrections to the $2\nu\beta\beta$ matrix element is
\begin{align}
	C_{2\nu}^{\rm BSM} &= 
	\frac{8 g_A^4}{3} 
    \Bigg\{ 
  (\tilde{\epsilon}_{R} - \tilde{ \epsilon}_L)^2  \left[ (M_{GT}^K)^2 + (M_{GT}^L)^2 +  M_{GT}^K  M_{GT}^L \frac{m_e^2}{E_{e1}E_{e2}} \right] \nonumber \\
    &+ \epsilon_T\frac{g_T}{g_A}  \left[(M_{GT}^K)^2 +  (M_{GT}^L)^2 +  M_{GT}^K M_{GT}^L\right]\, \frac{m_e (E_{e1} + E_{e2})}{E_{e1} E_{e2}} \Bigg\} \\
    & \simeq 8 g_A^4 \left(M_{GT}^{(-1)})\right)^2 
    \left( (\tilde{\epsilon}_{R} - \tilde{ \epsilon}_L)^2 \left(\frac{2}{3} + \frac{1}{3}\frac{m_e^2}{E_{e1}E_{e2}}  \right)
    + \epsilon_T\frac{g_T}{g_A} \frac{m_e (E_{e1} + E_{e2})}{E_{e1} E_{e2}}  
    \right)\,,
\end{align}
where we neglected higher-order terms in the $\epsilon_{K,L}$ expansion. 
Here the isovector tensor charge is \cite{Gupta:2018qil,FlavourLatticeAveragingGroupFLAG:2021npn}
\begin{eqnarray}
    g_T = 0.989(34)\,.
\end{eqnarray}

\begin{figure}[b]
    \centering
    \includegraphics[width=0.49\linewidth]{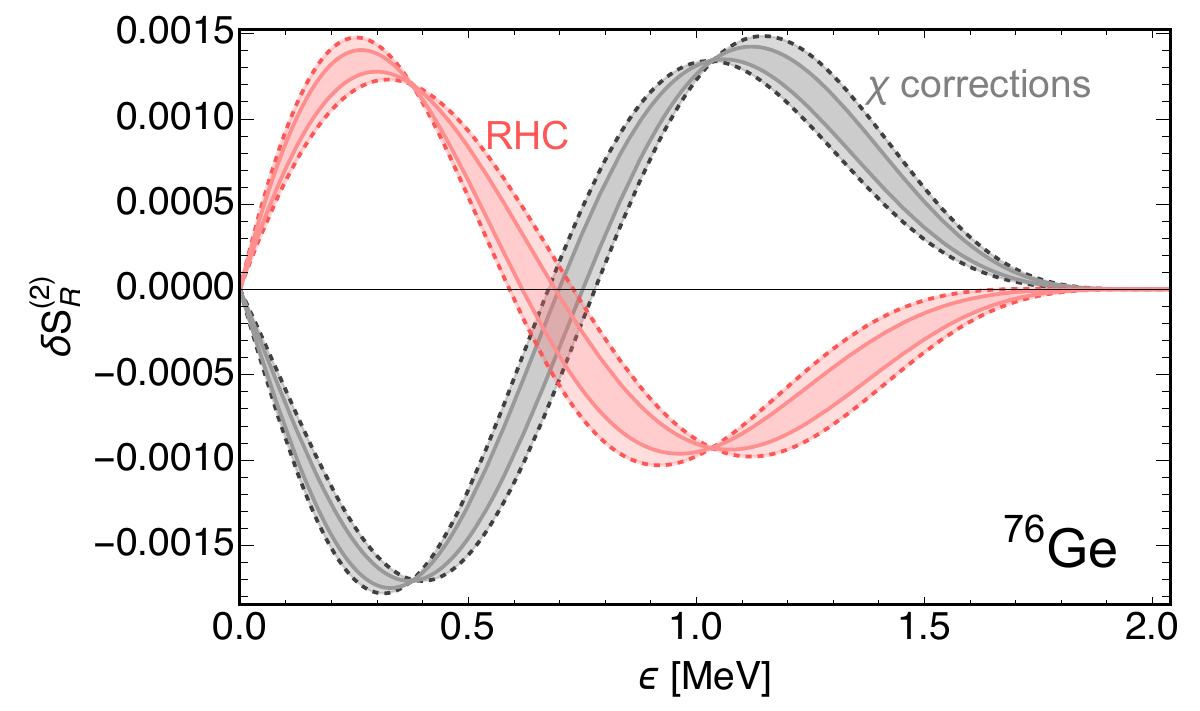}
    \includegraphics[width=0.49\linewidth]{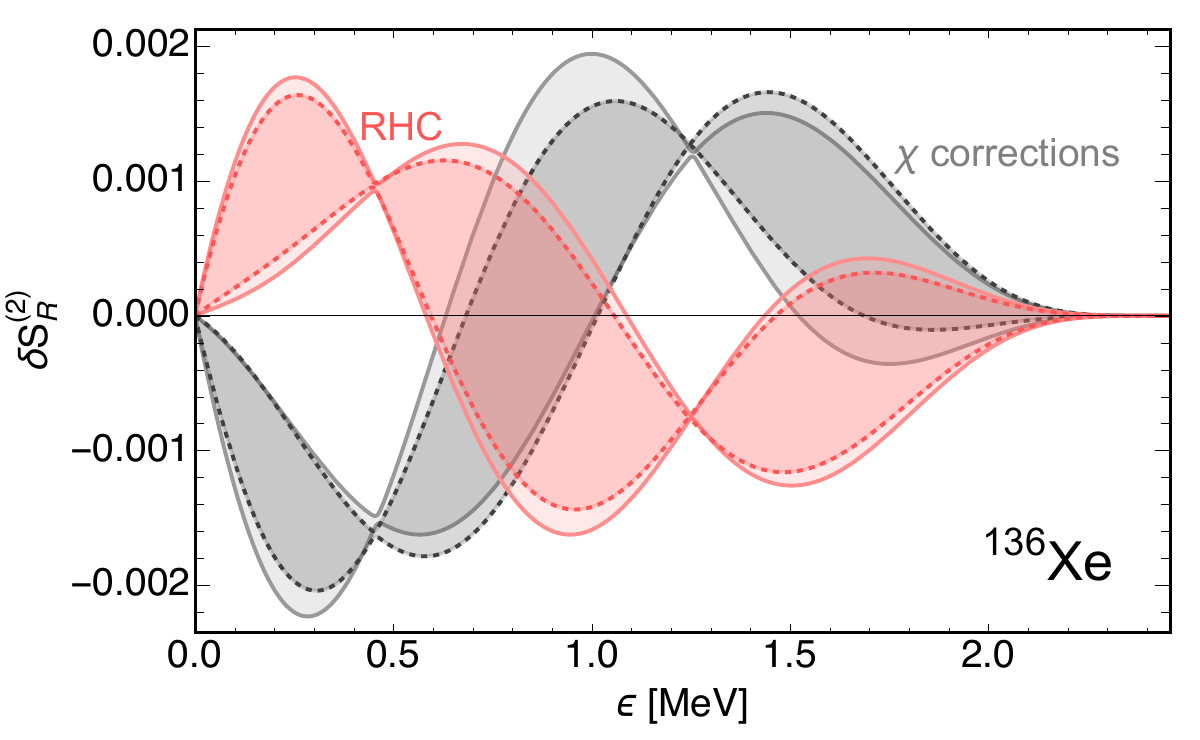}
    \caption{Shape-factor deviations $\delta S$ induced by the BSM right-handed vector current specified in Eq.~\eqref{eq:bsmlagrangian} for both Germanium (left) and Xenon (right). Here, we take $|\tilde{\epsilon}_L - \tilde\epsilon_R| =0.1$.}
    \label{fig:epsilonLR}
\end{figure}
Fig. \ref{fig:epsilonLR} shows the effect of adding a coupling of the nucleon to a leptonic right-handed charged-current (RHC) on the $2\nu\beta\beta$ decay of $^{76}$Ge (left) and $^{136}$Xe (right). 
We set $|\tilde \epsilon_L - \tilde \epsilon_R| = 0.1$, which is close to the bound in Eq. \eqref{eq:bounds}.
In red we show the distortion of the spectral shape induced by 
$|\tilde \epsilon_L - \tilde \epsilon_R|$,
compared to the corrections from WM and the pion-exchange two-neutrino potential, showed in gray. The bands denote the impact of the nuclear uncertainties, dominated by the uncertainty on $\xi_{31}$.  
Fig. \ref{fig:epsilonLR} shows that a measurement of the spectrum with $0.1\%$ accuracy could provide a competitive constraint on RHC, and, due to the different shape, the effect cannot be washed out by the nuclear uncertainty on $\xi_{31}$. The RHC contribution would be of the same size as the chiral corrections, which need to be included to obtain the correct bounds.

\begin{figure}[t]
    \centering
    \includegraphics[width=0.49\linewidth]{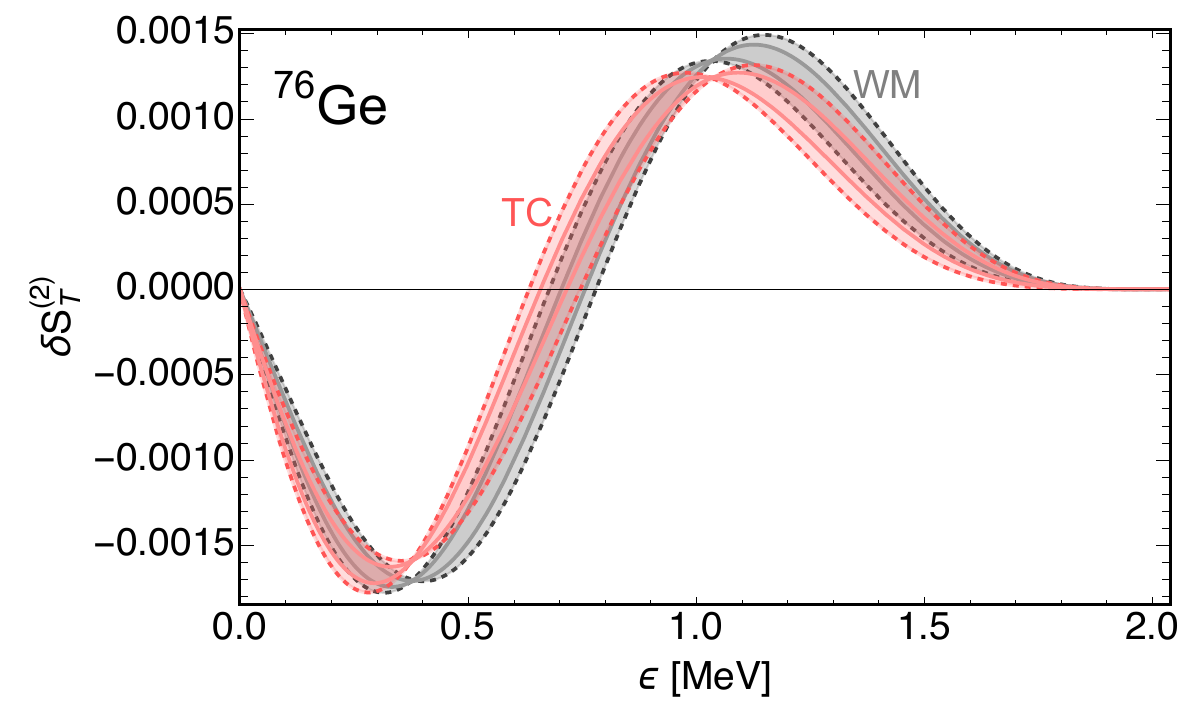}
        \includegraphics[width=0.49\linewidth]{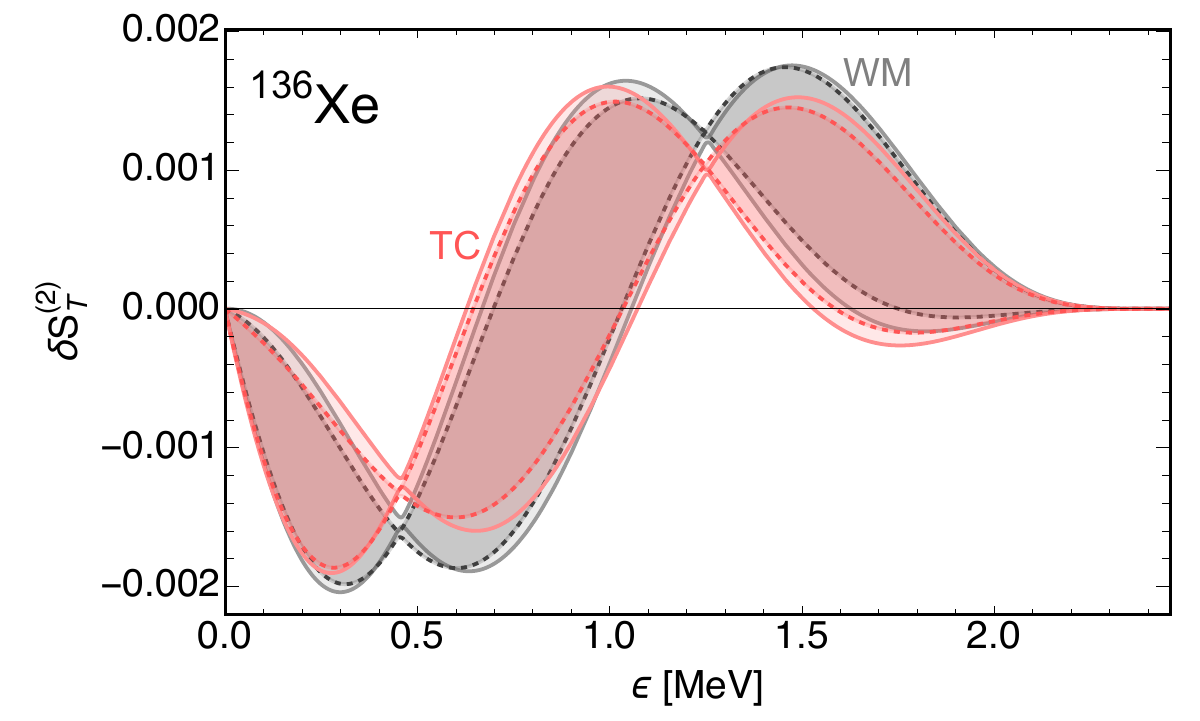}
    \caption{Shape-factor deviations $\delta S$ induced by the BSM tensor current (TC) with $\epsilon_T=-0.0014$ in comparison with the deviations caused by the weak magnetism (WM) correction term. }
    \label{fig:tensor2}
\end{figure}

Fig. \ref{fig:tensor2} shows the effect of a tensor coupling, set to $\epsilon_T =  -0.0014$. Also in this case, the effect cannot be mimicked by shifting the value of $\xi_{31}$, so that a $2\nu\beta\beta$ bound on tensor interactions would not suffer from nuclear theory uncertainties.
On the other hand, the tensor distortion has a shape that is very similar to the WM. In particular,  a tensor coupling of $\epsilon_T =  -0.0014$ would roughly double the shift due to WM, while $\epsilon_T =  +0.0014$ would cancel it. Searches for tensor couplings in the $2\nu\beta\beta$ spectrum can thus be competitive with Eq. \eqref{eq:bounds},
and with future searches the $\beta$ decay spectra of the neutron, $^6$He and $^{19}$Ne \cite{Brodeur:2023eul}, but they require the inclusion of chiral corrections, in addition to precise experiments.

\section{Conclusion}\label{sec:conclusions}

Next-generation $0\nu\beta\beta$ and dark-matter direct detection experiments will measure a record number 
of $2\nu\beta\beta$ events, 
resulting in high precision determination of the electron spectrum. 
Although often seen as a background, the $2\nu\beta\beta$ electron spectrum is interesting in its own right. In this work, we investigated the electron spectrum in the framework of chiral effective field theory. Our main results can be summarized 
as follows: 
\begin{itemize}
\item  We computed the leading corrections that appear at $\mathcal O(\mathcal Q/ \Lambda_\chi)$, 
where $\mathcal Q$ is the scale associated to the Q-value of the decay and $\Lambda_\chi \sim {\rm GeV}$ is 
the breakdown scale of chiral EFT. 
The leading corrections can be divided into three contributions. The first  arises from the multipole expansion of the weak currents, which receives a first contribution from weak magnetism. The second category consists of double-weak pion-exchange diagrams. The third category are short-range double-weak neutron-neutron-proton-proton-electron-electron-neutrino-neutrino interactions. 
\item  Computing the effect on the (differential) decay rate requires in principle complicated nuclear many-body calculations. 
This task is somewhat simplified by using 
the lepton energy expansion of Ref.~\cite{Simkovic:2018rdz}.
The corrections from weak magnetism then do not require new NMEs, while the pionic and short-distance effects do require NMEs but these are connected to the same NMEs necessary to describe $0\nu\beta\beta$. This is interesting as it implies that $0\nu\beta\beta$ NMEs can, in principle, be isolated in $2\nu\beta\beta$ measurements. 
Moreover, two currently unknown contact couplings are required for a full analysis and these may be obtained 
in the future by  Lattice QCD calculations~\cite{Shanahan:2017bgi,Tiburzi:2017iux}.

\item As a side result, we have performed a check of the convergence of the lepton energy expansion of Ref.~\cite{Simkovic:2018rdz} in case of the nuclear Shell Model using unexpanded calculations available in the literature. We conclude that the lepton energy expansion converges quickly and is a very useful tool. 

\item The chiral corrections increase the total decay rates of ${}^{76}$Ge and ${}^{136}$Xe by, respectively, $3\%$ and $10\%$, but these effects are swamped by the theoretical uncertainty of the total rate computation. We therefore focused on the normalized differential decay spectrum (the shape factor). Weak magnetism corrections are under good theoretical control and they shift the shape factor by a few per-mille over the entire spectrum. The corrections change the peak of the spectrum and can be isolated from nuclear structure corrections arising from subleading terms in the lepton energy expansion. This is in particular true around the nodes, special points in the spectrum where nuclear structure corrections vanish and where ideally the perturbation is maximal. 

\item The pionic corrections are small for ${}^{76}$Ge and at the few per-mille level for ${}^{136}$Xe, although they might become larger depending on the unknown contributions of the short-distance low-energy constants. They will be harder to identify because their effects are convoluted with nuclear structure corrections to the spectrum. In particular, the largest pionic corrections are proportional to $\xi_{31}$, a ratio of NMEs, that is poorly known. We have discussed how uncertainties in the NME ratios propagate to uncertainties in the shape factors and how this affects extracting the pionic contributions from data. With the present theoretical control of $\xi_{31}$, it is difficult 
to identify the pionic terms from measurements. That being said, a simultaneous fit of the $\xi$ ratios and the pionic corrections might be feasible in the future  and we encourage the experimental collaborations to pursue such a fit.

\item Precision $2\nu\beta\beta$ spectra measurements have been used to search for BSM physics. We demonstrated that the chiral corrections identified in this work should be included in order to avoid misidentifying BSM physics. In particular, weak magnetism closely resembles an exotic tensor interaction associated to BSM physics at an energy scale around $10$ TeV. In addition, per-mille measurements of the electron spectrum could set the most stringent limits on leptonic right-handed charged currents. 

\item Throughout our analysis of the $2 \nu \beta \beta$ spectrum's sensitivity to chiral correction and BSM physics, 
in contrast with most of the existing literature on the subject, 
we have taken a rather conservative / realistic approach and have systematically propagated the 
uncertainty due to nuclear matrix elements, such as the ratio $\xi_{31}$.

\end{itemize}

This first study of $2\nu \beta \beta$  decay using EFT methods has revealed that chiral corrections to the spectra 
are at the sub-percent level, while corrections to the decay rate can be at the 10\% level. In both cases, 
a full assessment of the impact requires better control of nuclear structure effects. 
These observations open up several future lines of investigation. 
On one hand, the size of chiral corrections to the spectra suggests that control over electromagnetic radiative corrections 
will ultimately be required to identify the chiral effects.  EFT methods have been shown to be beneficial in  
analyzing radiative corrections  to nuclear weak processes~\cite{Cirigliano:2024rfk,Cirigliano:2024msg,Hill:2023acw,Hill:2023bfh} and can be generalized to study  radiative corrections 
to double weak processes such as $2 \nu \beta \beta$.
On the other hand, the study of chiral corrections to the decay rate, in conjunction  with first-principles nuclear 
structure calculations, can lead in the future to precise  predictions for the  $2 \nu \beta \beta$ decay rates 
across a number of isotopes, thus increasing our confidence on the whole methodology used to study the 
corresponding neutrino-less decays.

\section*{Acknowledgements} 
We thank Maxime Pierre, Auke-Pieter Colijn, Javier Men\'endez and Ralph Massarczyk for useful discussions and encouragement. 
 Financial support by the Dutch Research Council (NWO) in the form of a VIDI grant, the U.S.\ DOE (Grant No.\
DE-FG02-00ER41132), and  Los Alamos
National Laboratory's Laboratory Directed Research and Development program under projects
20250164ER and 20230047DR 
is gratefully acknowledged. 
We also acknowledge support from the Dutch Research Council (NWO), under project number VI.Veni.222.318 and from Charles University through project PRIMUS/24/SCI/013.
Los Alamos National Laboratory is operated by Triad National Security, LLC,
for the National Nuclear Security Administration of U.S.\ Department of Energy (Contract No.\
89233218CNA000001). 
We acknowledge support from the DOE Topical Collaboration ``Nuclear Theory for New Physics,'' award No.\ DE-SC0023663. 
This material is based, in part, on work supported by the U.S. Department of Energy, Office of Science, Office of Workforce Development for Teachers and Scientists, Office of Science Graduate Student Research (SCGSR) program, under contract number DE‐SC0014664.

\bibliographystyle{utphysmod}
\bibliography{bibliography}

\end{document}